\documentclass[11pt,a4paper]{article}
\usepackage{jheppub}
\usepackage{cases}
\usepackage{graphicx}
\usepackage{dcolumn,color}
\usepackage{bm}
\usepackage{epstopdf}
\usepackage{epsfig,subfigure}
\usepackage{color,framed}
\usepackage{amsmath,amsthm,amssymb} 

\usepackage[table]{xcolor}

\title{\boldmath Gravitational resonances in mimetic thick branes}

\author[a,b]{Yi Zhong,}
\author[a]{Yu-Peng Zhang,}
\author[a]{Wen-Di Guo}
\author[a,c,1]{and Yu-Xiao Liu,\note{Corresponding author.}}

\affiliation[a]{Institute of Theoretical Physics $\&$ Research Center of Gravitation,
            Lanzhou University,\\
            Lanzhou 730000, P. R. China}
\affiliation[b]{ School of Physics and Electronics Science, Hunan University,\\
            Changsha 410082, P. R. China}
\affiliation[c]{Key Laboratory for Magnetism and Magnetic of the Ministry of Education, Lanzhou University,\\
            Lanzhou 730000, P. R. China}

\emailAdd{zhongy@hnu.edu.cn}
\emailAdd{zhangyupeng14@lzu.edu.cn}
\emailAdd{guowd16@lzu.edu.cn}
\emailAdd{liuyx@lzu.edu.cn}

\abstract{In this work, we investigate gravitational resonances in both single and double mimetic thick branes, which can provide a new way to detect the extra dimension. For the single brane model, we apply the relative probability proposed in [Phys. Rev. D. 80 (2009) 065019].  For the double brane model, we investigate the resonances quasi-localized on the double brane, on the
sub-branes and between the sub-branes, respectively. To investigate the resonances quasi-localized on the double brane, we introduce two different definitions of the relative probability and find that the corresponding mass spectra of gravitational resonances are almost the same. For the gravitational resonances quasi-localized on sub-branes and between the sub-branes, the influence of the distance between the two sub-branes and the thickness of the sub-branes are analyzed and new features are found in both cases.}

\begin{document}
\maketitle

\section{Introduction}
The braneworld scenario has attracted much attention since the renowned Arkani-Hamed-Dimopoulos-Dvali
(ADD) model and Randall-Sundrum (RS) model were proposed \cite{ArkaniHamed:1998rs,Randall:1999ee,Randall:1999vf}. It is possible to solve the hierarchy problem and the cosmological constant problem in the braneworld scenario \cite{ArkaniHamed:1998rs,Randall:1999ee,Randall:1999vf,Kim:2000mc}. Both the ADD model and RS model are thin brane models. Later, various thick brane models were investigated \cite{DeWolfe:1999cp,Gremm:1999pj,Csaki:2000fc,Kobayashi:2001jd,Giovannini:2001fh,Bazeia:2003aw,Bazeia:2004dh,Gremm:2000dj,Bazeia:2002sd,Afonso:2006gi,Dzhunushaliev:2008gk,Neupane:2010ey} and the localization of matter fields on the brane was realized. In the braneworld scenario, our four-dimensional universe is an infinitely  thin brane or a domain wall embedded in a higher dimensional space-time. The Standard Model fields are localized on the brane \cite{Bajc:1999mh,Ringeval:2001cq,Bagger:2004rr,Liu:2007ku,Liu:2008wd,Ghoroku:2001zu,Kehagias:2000au}, while the gravity propagates in all dimensions. In order not to contradict the present observations, the zero mode of the tensor perturbation of gravity should be localized on the brane and recover the four-dimensional Newtonian potential \cite{Randall:1999vf,Csaki:2000fc}. In many types of brane models, the extra part of the tensor perturbation satisfies a Schr\"{o}dinger-like equation, and the effective potential may support resonance modes \cite{Csaki:2000fc,Csaki:2000pp,Gremm:1999pj,Cruz:2013uwa,Xie:2013rka,Xu:2014jda,Yu:2015wma}.
Thus, apart from the localized zero mode, the quasi-localized modes, i.e. the resonance modes may exist and contribute correction to the four-dimensional Newtonian potential \cite{Csaki:2000fc,Csaki:2000pp,Gremm:1999pj,Cruz:2013uwa,Xu:2014jda}, which can provide a new way to detect the extra dimension. Therefore, the investigation of gravitational resonances is an important topic in braneworld models. Gravitational resonances also appear in other systems, e.g., the quasi-normal modes outside of black holes \cite{Nollert:1999ji}. For a recent review, see Ref.~\cite{Liu:2017gcn}.

Since it is widely believed that general relativity should be modified, which is inspired by both the theoretical motivation and cosmological observation data \cite{Nollert:1999ji,Sotiriou:2008rp,Nojiri:2006ri,Nojiri:2017ncd}, braneworld models in modified gravities were investigated  extensively \cite{Kastor:2017knv,Bazeia:2017mjd,Afonso:2007zz,Guo:2014bxa,German:2013sk,Arias:2002ew,
Zhong:2015pta,Yang:2012hu,daSilva:2017jbx,Yang:2017evd,Guo:2015qbt,Guo:2018tpo,Karam:2018squ}. New features such as inner structure of branes and pure geometrical branes were found \cite{Arias:2002ew,Zhong:2015pta,Yang:2012hu}.
Recently the mimetic gravity was proposed to solve the dark matter problem \cite{Chamseddine:2013kea,Chamseddine:2014vna}. In this theory, the
physical metric $g_{\mu\nu}$ is defined in terms of an auxiliary metric $\hat{g}_{\mu\nu}$
and a scalar field $\phi$ as
$g_{\mu\nu}=-\hat{g}_{\mu\nu}\hat{g}^{\alpha\beta}\partial_\alpha\phi\partial_\beta\phi$. In the framework of mimetic gravity, the geometrical explanation of dark matter was given on galaxy level, cluster level and cosmological evolution and perturbation level \cite{Dutta:2017fjw,Matsumoto:2015wja,Myrzakulov:2015kda,Vagnozzi:2017ilo}. It is also possible to unify the inflation and late-time acceleration period in this theory \cite{Nojiri:2014zqa}. For more recent works of mimetic gravity in cosmology see Refs. \cite{Odintsov:2015ocy,Odintsov:2016imq,Odintsov:2016oyz,
Nojiri:2016vhu,Odintsov:2018ggm}. The Friedmann-Robertson-Walker thin brane was considered and the late time cosmic expansion was explained in the favor of observational data, and the initial time cosmological inflation was also produced \cite{Sadeghnezhad:2017hmr}. Later, thick branes with inner structure generated by mimetic scalar field were found in Ref. \cite{Zhong:2017uhn}. The gravitational perturbation was analyzed in detail. It was found that the tensor zero mode is localized on the branes. For specific parameters, the branes split into multi sub-branes, and the effective potential of the tensor perturbation also splits into multi-wells, which is different from the usual braneworld case. Unlike the zero mode of the tensor perturbation which is localized on the sub-brane, the resonance modes can be quasi-localized on or between the sub-branes. This is an important new feature in the mimetic thick brane. Inspired by this we would like to study the resonances of the tensor perturbation in these brane models. We will introduce alternative definitions of relative probability and compare the corresponding mass spectra of gravitational resonances. Then we will analyze how the structure of the brane impacts on the gravitational resonances quasi-localized on different locations of the double brane.

The organization of this paper is as follows. In Sec.~\ref{Sec2}, we briefly introduce the mimetic thick brane model and the tensor perturbation of gravity. In Sec.~\ref{Sec3}, we investigate the gravitational resonances in both single and double mimetic brane models. In Sec.~\ref{Sec4}, we will discuss the contribution of the resonances to the four-dimensional Newtonian potential and give a conclusion.

\section{Linear perturbation in a mimetic thick brane model}\label{Sec2}

In this section, we briefly introduce the mimetic thick brane model and the linear  perturbation of the brane system, which were given in Ref. \cite{Zhong:2017uhn} in detail. We take the geometrized units in which the gravitational constant $\kappa^2=1$. The action of the five-dimensional mimetic gravity is
    \begin{eqnarray}
        S=\int d^5x \sqrt{-g} \left( \frac{R}{2}
        + \lambda\left[\partial^M \phi \partial_N \phi-U(\phi)\right]-V(\phi) \right),  
        \label{action mgb1}
    \end{eqnarray}
where $\lambda$ is a Lagrange multiplier. Throughout this paper, the indices $M,N\cdots=0,1,2,3,5$ denote the bulk coordinates and $\mu,\nu\cdots=0,1,2,3$ denote the ones on the brane.

 In the original mimetic gravity, $U(\phi)=-1$ , and $\phi$ represents the conformal degree of the metric $g_{MN}$ \cite{Chamseddine:2013kea}. This theory was extended to $U(\phi)<0$ for cosmological application \cite{Astashenok2015}. In this work, the mimetic scalar field $\phi=\phi(y)$ is used to generate the thick brane. Therefore, we assume that $U(\phi)=g^{MN}\partial_M \phi \partial_N \phi>0$ \cite{Zhong:2017uhn}. This generalization can also provide thick branes with inner structure.
 The equations of motion can be easily obtained by varying the action with respect to the physical metric $g_{MN}$, the mimetic scalar field $\phi$ and the Lagrange multiplier $\lambda$ \cite{Zhong:2017uhn}:
    \begin{eqnarray}
        \label{var eom1}
        G_{MN}+2\lambda \partial_M \phi \partial_N \phi-L_{\phi}g_{MN}=0,    \\
        \label{var eom2}
        2\lambda\Box^{(5)}\phi+2\nabla_{M}\lambda\nabla^{M}\phi+\lambda \frac{\partial U}{\partial \phi}+\frac{\partial V}{\partial \phi}=0, \\
        \label{var eom3}
        g^{MN}\partial_M \phi \partial_N \phi-U(\phi)=0.
    \end{eqnarray}
 {where $
L_{\phi}=\lambda\left[g^{MN}\partial_M \phi \partial_N \phi-U(\phi)\right]-V(\phi)$
is the lagrangian of the mimetic scalar field.}
 Assuming the Minkowski brane metric
    \begin{eqnarray}
        \label{brane metric1}
       ds^2=a^2(y)\eta_{\mu\nu}dx^{\mu}dx^{\nu}+dy^2,
    \end{eqnarray}
the {equations of motion} (\ref{var eom1})-(\ref{var eom3}) read
    \begin{eqnarray}
        \label{eom21}
      \frac{3a'^2}{a^2}+\frac{3a''}{a}+V(\phi)+\lambda \left(U(\phi)-\phi'^2\right)&=&0, \\
       \label{eom22}
       \frac{6a'^2}{a^2}+V(\phi)+2\lambda \left(U(\phi)+\phi'^2\right)&=&0, \\
       \label{eom23}
       \lambda \left(\frac{8a'\phi'}{a}+2\phi''+\frac{\partial U}{\partial \phi}\right)
        +2\lambda'\phi'+\frac{\partial V}{\partial \phi}&=&0,  \\
       \label{eom24}
       \phi'^2-U(\phi)&=&0,
    \end{eqnarray}
where the primes denote the derivatives with respect to the extra-dimensional coordinate $y$. Since there are only three independent equations in Eqs. (\ref{eom21})-(\ref{eom24}) and five independent variables, we can easily solve $\lambda(y)$, $V(\phi)$ and $U(\phi)$ for any given $a(y)$ and $\phi(y)$. As we will see later, the equation of the tensor perturbation depends only on the warped factor. Therefore, in the next section we will only give the warped factor $a(y)$ and omit the expression of $\phi(y)$, $\lambda(y)$, $V(\phi)$ and $U(\phi)$ for the brane models.
When the mimetic scalar $\phi(y)$ is set to be constant, the potentials $V(\phi)$ and $U(\phi)$ and the Lagrange multiplier $\lambda(y)$ also become constants, and the theory turns to general relativity. In this work we will focus on the double brane models, in which $\phi(y)$ can not be a constant.

Next, we consider the linear perturbation of the brane system. It is easy to see that the scalar, vector
and tensor modes of the pertubation are decoupled with each other. Thus, we will investigate the tensor and scalar perturbation separately.
Redefining the extra-dimensional coordinate $ dz=\frac{1}{a(y)}dy $, the perturbed metric in the new coordinate is given by
    \begin{eqnarray}
        \label{tensor metric}
        ds^2=a(z)^2[(\eta_{\mu\nu}+h_{\mu\nu})dx^{\mu}dx^{\nu}+dz^2],
    \end{eqnarray}
where the tensor perturbation $h_{\mu\nu}=h_{\mu\nu}(x^{\mu},z)$ depends on all the coordinates and satisfies the transverse-traceless (TT) condition $\eta^{\mu\nu}\partial_\mu h_{\lambda\nu}=0$ and $\eta^{\mu\nu}h_{\mu\nu}=0$. Next we redefine the perturbation as
$h_{\mu\nu}=a(z)^{-\frac{3}{2}}\tilde{h}_{\mu\nu}$. After tedious but straightforward derivation, the perturbation of Eq. (\ref{var eom1}) yields
    \begin{eqnarray}
        \Box^{(4)}\tilde{h}_{\mu\nu}+\partial^2_{z}\tilde{h}_{\mu\nu}
        -\frac{\partial^2_{z}a^{\frac{3}{2}}}{a^{\frac{3}{2}}}\tilde{h}_{\mu\nu}=0.
    \end{eqnarray}
Employing the decomposition $\tilde{h}_{\mu\nu}=\epsilon_{\mu\nu}(x^\gamma) \text{e}^{ip_{\lambda}x^{\lambda}}t(z)$ with the polarization tensor $\epsilon_{\mu\nu}$ satisfying the TT condition $\eta^{\mu\nu}\partial_\mu \epsilon_{\lambda\nu}=0$ and $\eta^{\mu\nu}\epsilon_{\mu\nu}=0$,
 one can obtain the Schr\"{o}dinger-like equation for the extra part $t(z)$ of the tensor perturbation:
    \begin{eqnarray}
        \label{eq tensor}
        -\partial^2_{z}t(z)+V_\text{t}(z)t(z)=m^2 t(z),
    \end{eqnarray}
where the effective potential $V_\text{t}(z)$ is given by
    \begin{eqnarray}
        V_\text{t}(z)=\frac{\partial^2_{z}a^{\frac{3}{2}}}{a^{\frac{3}{2}}}, \label{PotentialV}
    \end{eqnarray}
and $m$ is the mass of the tensor perturbation $t(z)$.
The zero mode of the tensor perturbation is
    \begin{eqnarray}
        t_0 (z)\propto a^{\frac{3}{2}}(z).
    \end{eqnarray}
Obviously, the tensor perturbation in mimetic gravity is the same as that in general relativity.
Furthermore, Eq. (\ref{eq tensor}) can be factorized as
    \begin{eqnarray}
        \label{eq tensor2}
        \left(-\partial_{z}+ \partial_{z} \ln a^{\frac{3}{2}} \right)
        \left(\partial_{z}+\partial_{z} \ln a^{\frac{3}{2}} \right)t(z)
        =m_t^2 t(z).
    \end{eqnarray}
The structure in the above equation ($\mathcal{K}\mathcal{K}^\dag$ with $\mathcal{K} = -\partial_{z}+ \partial_{z} \ln a^{\frac{3}{2}}$) ensures that the eigenvalues are non-negative and so there is no tensor tachyon mode with $m^2<0$. Thus, the brane is stable against the tensor perturbation \cite{Zhong:2017uhn}. The tensor zero mode is localized around the thick brane embedding in an AdS$_5$ space-time. Nevertheless, the mimetic scalar field generates more types of thick brane, which could lead to new type of potential of the tensor perturbation. Thus, one can expect new phenomena in the resonances of the tensor perturbation.

At last, we turn to the scalar perturbation.
The perturbed metric is
    \begin{eqnarray}
        \label{brane metric}
       ds^2=a^{2}(z)\left[(1+2\psi)\eta_{\mu\nu}dx^{\mu}dx^{\nu}+(1+2\Phi)dz^2\right],
    \end{eqnarray}
and the perturbed scalar field is $\phi+\delta\phi$. The field equations of the scalar perturbation $\delta\phi$ and the scalar modes $\Phi$ and $\psi$  are
    \begin{eqnarray}
       &&
    -\frac{3}{2}\partial^2_z \delta\phi
         +\frac{3}{4}\left(\frac{a^2}{\partial_z \phi}\frac{\partial U}{\partial \phi}+\frac{2\partial^2_z \phi}{\partial_z \phi}-\frac{4\partial_z a}{a}\right)\partial_z \delta \phi    \nonumber\\
        && +\left[\frac{3a\partial_z a}{\partial_z \phi}\frac{\partial U}{\partial \phi}+2\lambda(\partial_z \phi)^2+\frac{3}{4}a^2\left(\frac{\partial^2 U}{\partial \phi^2}-2\frac{\partial U}{\partial \phi}\frac{\partial^2_z \phi}{(\partial_z \phi)^2}\right)\right]\delta\phi = 0, \label{scalar master} \\
       && \Phi = -2\psi,	\label{ptb5} \\
       && \Phi = \frac{\partial_z \delta\phi}{\partial_z \phi}-\frac{a^2}{2(\partial_z \phi)^2}\frac{\partial U}{\partial \phi}\delta\phi.
    \end{eqnarray}
Redefining the perturbation of the scalar field as $\delta\phi(x^{\mu},z)=\frac{(\partial_z \phi)^{\frac{3}{2}}}{a^2}s(z)\overline{\delta\phi}(x^\mu)$,  we can obtain the equation of the extra part of the scalar perturbation:
     \begin{eqnarray}
        \label{scalar pertb}
        -\partial^2_z s(z)+V_\text{s}(z)s(z)=0,
    \end{eqnarray}
 where the effective potential $V_\text{s}(z)$ is given by
    \begin{eqnarray}
    	\label{scalar potential}
        V_\text{s}(z)=\frac{2(\partial_z a)^2-a\partial^2_z a}{a^2}
        +\frac{-(\partial^2_z \phi)^2+2\partial_z \phi \partial^{3}_z\phi}{4(\partial_z\phi)^2}.
    \end{eqnarray}
Since there is no term of the form $\Box^{4} \delta\phi$ in Eq. (\ref{scalar master}),  the scalar perturbation does not propagate on the brane.  Furthermore, for the following brane models, we will plot the shape of the potential $V_\text{s}(z)$ to show that the scalar perturbation is not localized on the brane, and thus does not contribute to the four-dimensional Newton potential.
{ Therefore, though it seems strange that there is no term of the form $\Box^{4} \delta\phi$ in Eq. (\ref{scalar master}), it does not lead to any problems. Similarly, in the cosmological context, the scalar perturbation of mimetic gravity has also no terms of  $\nabla^{2} \delta \phi$ (see Eq. (64) of Ref. \cite{Chamseddine:2014vna}) , which implies that the sound speed is identically zero.  }

\section{Gravitational resonance in various thick brane models}\label{Sec3}

In the above section, it was pointed out that the zero mode of the tensor perturbation is localized around the brane embedding in an AdS$_5$ space-time. In this section, we will investigate quasi-localized modes, i.e. the gravitational resonances, in both single and double brane models.

\subsection{Gravitational resonances in a single-brane model}\label{subSec31}
\begin{figure}[!htb]
    \begin{center}
    \subfigure[The warped factor]{
        \includegraphics[width=4cm]{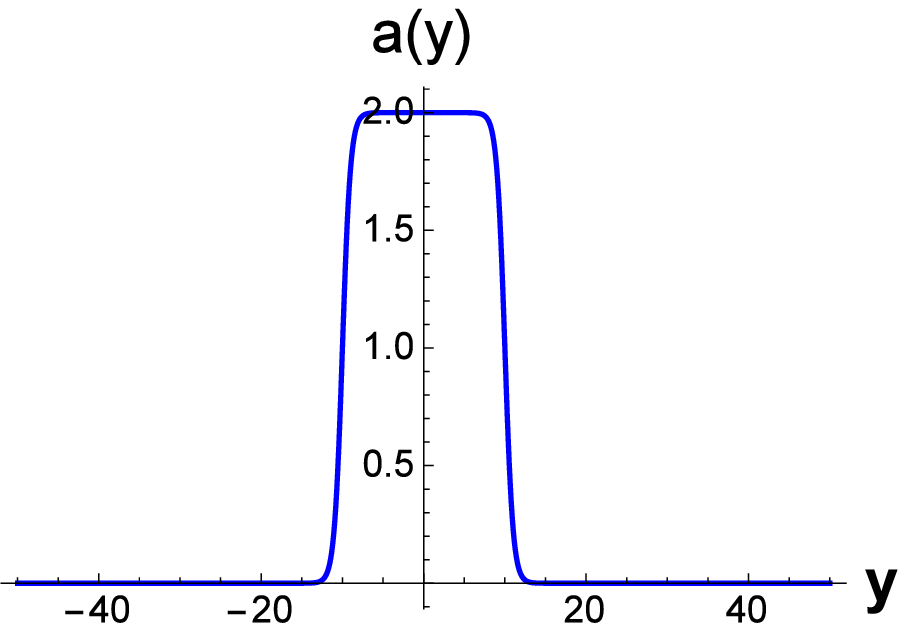}}
    \subfigure[The effective potential $V_\text{t} (z)$]{
        \includegraphics[width=4cm]{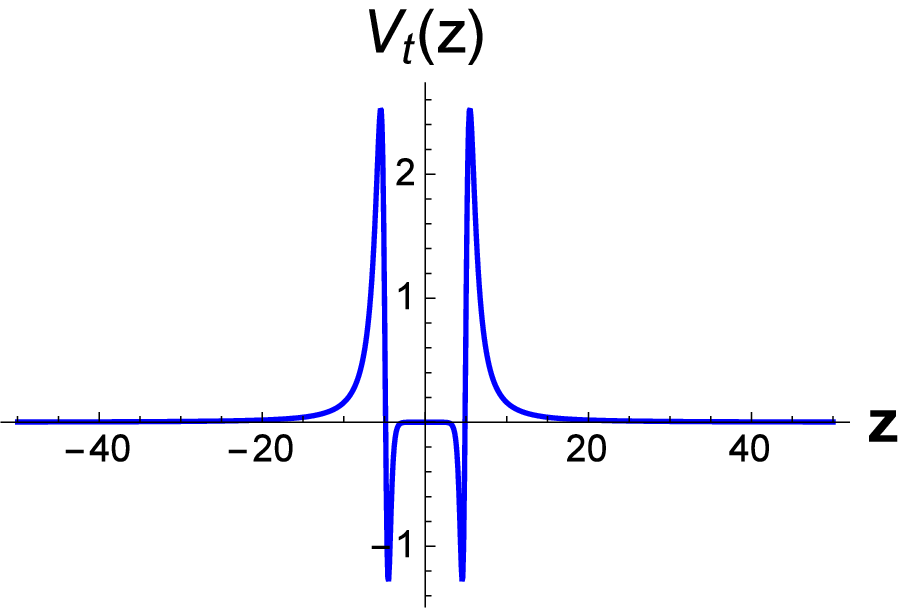}}
    \subfigure[ The effective potential $V_\text{s} (z)$]{
        \includegraphics[width=4cm]{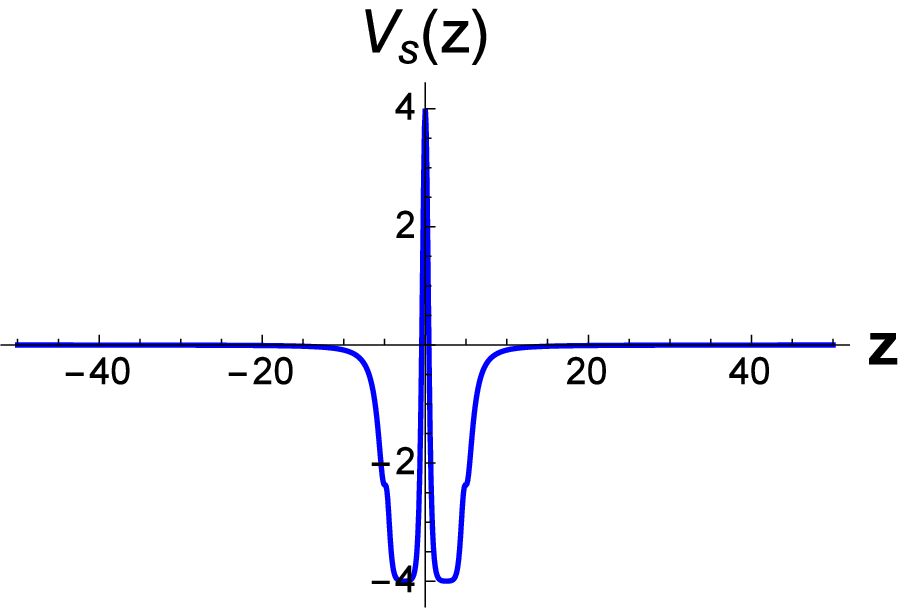}}
    \end{center}
    \caption{The shapes of the warp factor $a(y)$ and the effective potentials $V_\text{t} (z),~V_\text{s} (z)$ for the single brane model (\ref{WarpedFactorSingleBrane}). The parameters are set to $k=1$
             and $b=10$.} \label{figure 11}
    \end{figure}

First of all, as a simple example, we study a single brane model with the following warped factor \cite{Zhong:2017uhn}
    \begin{eqnarray}
        a(y) =\text{tanh}[k(y+b)] - \text{tanh}[k(y-b)]. \label{WarpedFactorSingleBrane}
    \end{eqnarray}
The shapes of the warped factor and the corresponding effective potentials (\ref{PotentialV}) and
(\ref{scalar potential}) are plotted in Fig.~\ref{figure 11}. It can be seen that the potential $V_\text{t}(z)$ has an obvious double-well with two barriers, which is the main reason leading to resonance KK modes.
The potential $V_\text{s}(z)$ approaches $0^-$ at infinity, therefore the scalar mode is not localized on the brane and does not contribute to the four-dimensional Newton potential.

Due to the complexity of the function $a(y)$ and the coordinate transformation, we can not obtain the analytical expressions of the warp factor $a(z)$ and the effective potential $V_\text{t}(z)$.
To solve the Schr\"{o}dinger-like equation (\ref{eq tensor}) for $t(z)$ numerically, we decompose $t(z)$ into  an even parity mode $t_\text{e}(z)$ and an odd parity mode $t_\text{o}(z)$, which  are set to satisfy the following boundary conditions:
    \begin{eqnarray}
        \label{condition even}
        t_\text{e} (0)=1,~~~~~~~~\partial_z t_\text{e} (0)=0;\\
        \label{condition odd}
        t_\text{o} (0)=0,~~~~~~~~\partial_z t_\text{o} (0)=1.
    \end{eqnarray}
To investigate the gravitational resonances, we adopt the concept of the relative probability of the KK mode $t(z)$ with mass $m$, which was defined in Ref.~\cite{Liu:2009ve,Almeida:2009jc}:
    \begin{eqnarray}
        \label{relative p1}
        P(m^2)=\frac{\int^{z_b}_{-z_b}|t(z)|^2 dz}{\int^{z_{max}}_{-z_{max}}|t(z)|^2 dz}.
    \end{eqnarray}
Here $2z_b$ is approximately the width of the thick brane, and $z_{max}=10z_b$.
Note that there are other methods that can find out KK resonances, such as the transfer matrix method \cite{DuZhao2013,Landim1105.5573}.

\begin{figure}[!htb]
    \begin{center}
    {\label{figure P 1}
        \includegraphics[width=4cm]{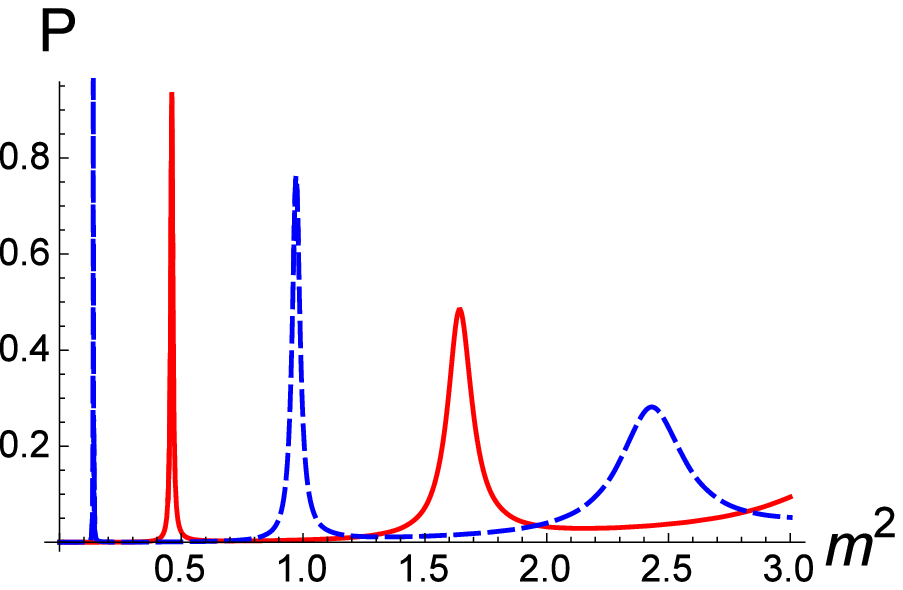}}
    \end{center}
    \caption{The relative probability $P(m^2)$ of the even parity mode $t_\text{e}$ (red solid lines) and the odd parity mode $t_\text{o}(z)$ (blue dashed lines) for the single brane model (\ref{WarpedFactorSingleBrane}). The parameters are set to $k=1$
             and $b=10$. } \label{figure 12}
    \end{figure}

For a given $m^2$, the
 Schr\"{o}dinger-like equation (\ref{eq tensor}) can be solved numerically for the even parity mode
  $t_\text{e}(z)$ and the odd parity mode $t_\text{o}(z)$ with the conditions (\ref{condition even}) and (\ref{condition odd}),
  respectively. Then the relative probability $P$ corresponding to this $t_\text{e}(z)$ or $t_\text{o}(z)$ can be obtained.
   By this means, the relative probability as a function of $m^2$ is obtained and plotted in
   Fig.~\ref{figure 12}, in which each of the peaks represents a resonance mode.

    \begin{table}
        \centering
        \caption{The mass spectrum $m_{n}^2$, relative probability $P_{max}$, FWHM $\Gamma$ and life-time $\tau$ of the resonances for the single brane model (\ref{WarpedFactorSingleBrane}). }
    \begin{tabular}{|c|c|c|c|c|c|}
  \hline
  $n$ & Parity & $m_{n}^2$ & $P_{max}$ & $\Gamma$ & $\tau$ \\
  \hline
 1 & odd & $0.1385$ &$0.9902$ & $0.002418$ & $413.50$ \\
  \hline
 2 & even & $0.4609$ &$0.9367$ & $0.007880$ & $126.90$ \\
  \hline
 3 &   odd & $0.9711$ &$0.7609$ & $0.019633$ & $50.93$ \\
  \hline
 4 & even & $1.6428$ &$0.4852$ & $0.047884$ & $20.884$ \\
  \hline
 5 & odd & $2.4317$ &$0.2817$ & $0.107172$ & $9.3308$ \\
  \hline
\end{tabular}
        \label{table p1}
    \end{table}

 Furthermore, the corresponding life-time $\tau$ of
 the resonances can obtained by $\tau=\frac{1}{\Gamma}$, where $\Gamma$ is the full width at half
 maximum (FWHM) \cite{Gregory:2000,Almeida:2009jc}. The resonances having large life-time can be quasi-localized on the brane for a long time. Therefore, these resonances are approximately four-dimensional gravitons \cite{Csaki:2000pp}. The mass spectrum, FWHM, and life-times are shown in Table
 \ref{table p1}. It is shown that the relative probability $P$ and life-time $\tau$ of the
 resonance modes decrease with the mass square $m^2$, while the FWHM $\Gamma$ increases with $m^2$. Thus, the behavior of the resonances in the single mimetic brane is similar to that in a single brane model in general relativity \cite{Csaki:2000fc,Xie:2013rka}.

\begin{figure}[!htb]
    \begin{center}
    \subfigure[$m^2=0.1385$]{\label{figure odd 1}
        \includegraphics[width=4cm]{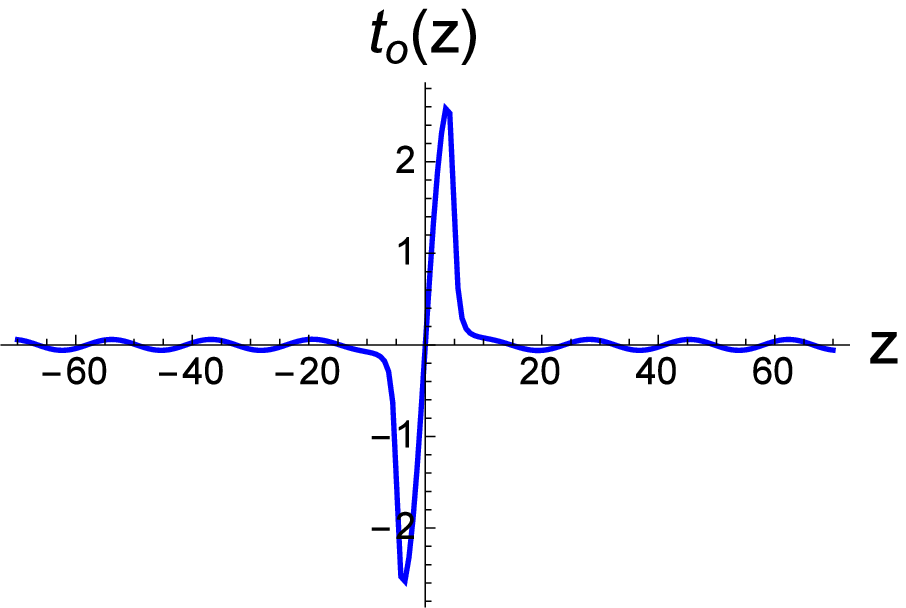}}
    \subfigure[$m^2=0.4609$]{\label{figure even 1}
        \includegraphics[width=4cm]{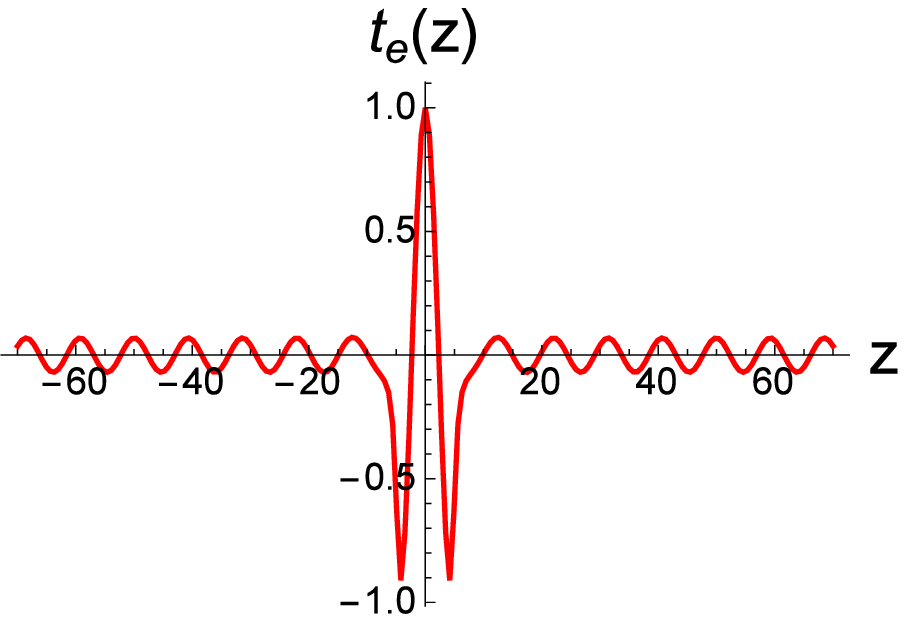}}
    \end{center}
    \caption{Plots of the first even and fist odd parity resonance modes for the single brane model (\ref{WarpedFactorSingleBrane}). The parameters are set to $k=1$, $b=10$, and $m^2=0.1385$ for $t_\text{o} (z)$, and $m^2=0.4609$ for $t_\text{e} (z)$.} \label{figure 13}
    \end{figure}

The wave functions of the odd and even modes corresponding to the highest peaks in Fig.~\ref{figure 12} are
plotted in Fig.~\ref{figure 13}.

\subsection{Gravitational resonances quasi-localized on a double-brane}\label{subSec32}
\label{sbs 32}
\begin{figure}[!htb]
    \begin{center}
    \subfigure[The warped
    factor]{\label{figure warped factor 2}
        \includegraphics[width=4cm]{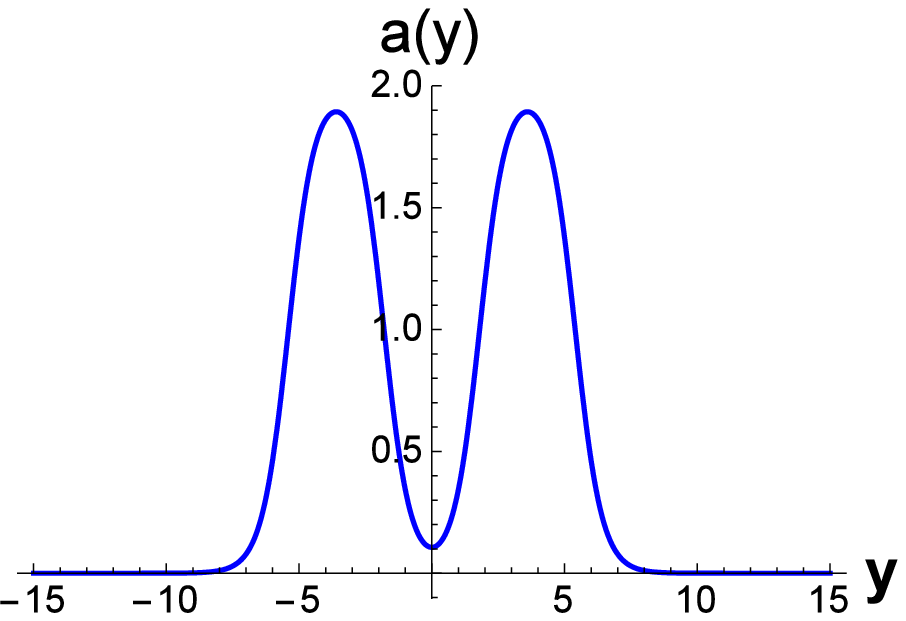}}
    \subfigure[The effective potential $V_\text{t} (z)$]{\label{figure potential 2}
        \includegraphics[width=4cm]{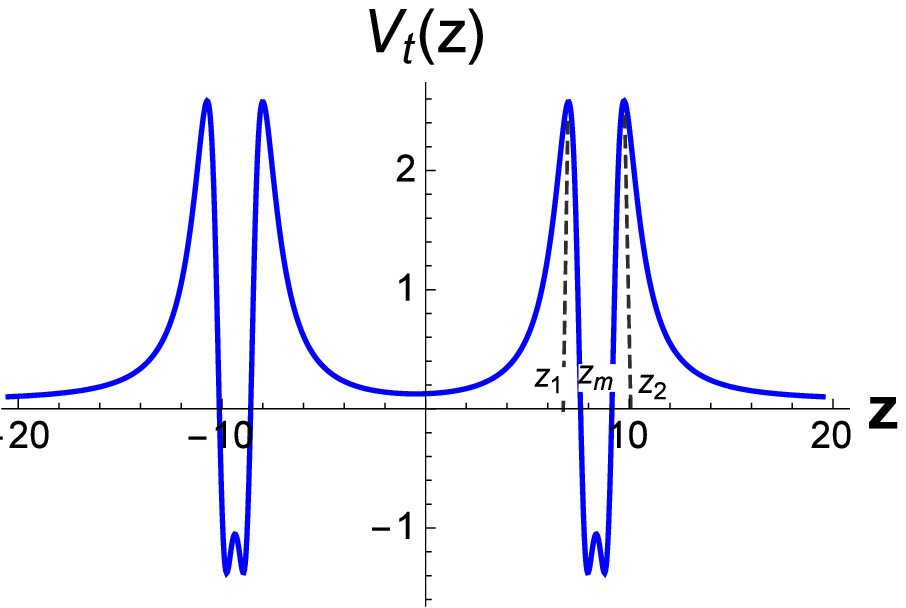}}
    \subfigure[The effective potential $V_\text{s} (z)$]{\label{figure potential 2}
        \includegraphics[width=4cm]{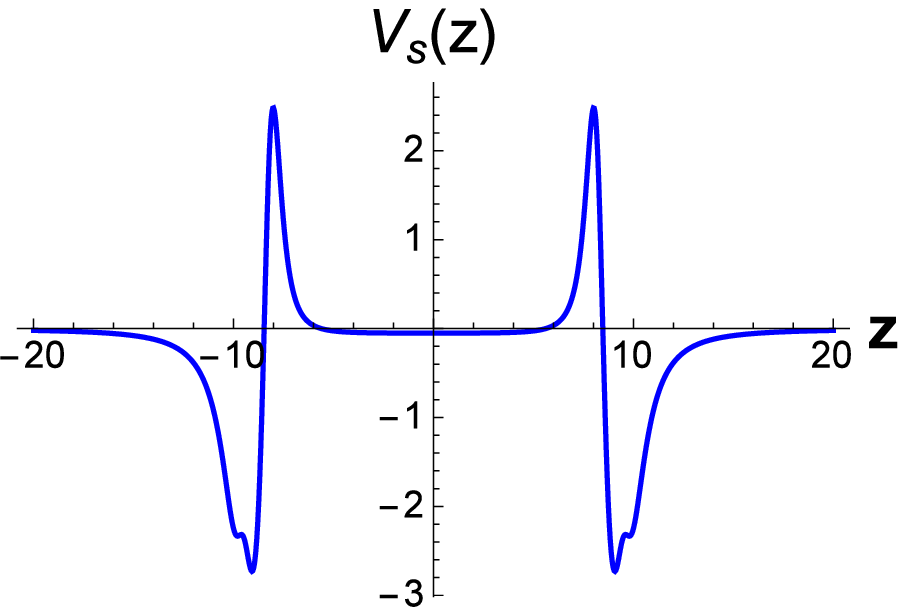}}
    \end{center}
    \caption{The warped factor $a(y)$ and the effective potentials $V_\text{t} (z)$ and $V_\text{s} (z)$ for the double brane model (\ref{WarpedFactorDoubleBrane}). The parameters are set to $k=1$,
             and $b=1.8$.} \label{figure 21}
    \end{figure}

Next we study the gravitational resonances quasi-localized on a double-brane. The warped factor is given by \cite{Zhong:2017uhn}
    \begin{eqnarray}
          a(y) &=& \text{tanh}[k(y+3b)]-\text{tanh}[k(y-3b)] \nonumber \\
               &-& \text{tanh}[k(y+b)]+\text{tanh}[k(y-b)].  \label{WarpedFactorDoubleBrane}
    \end{eqnarray}
The shapes of the warped
factor (\ref{WarpedFactorDoubleBrane}) and the corresponding effective potential (\ref{PotentialV}) are plotted in Fig.~\ref{figure 21},
which shows that the effective potential  $V_\text{t}(z)$ has two sub-wells, and the part between the two sub-wells can
also be regarded as a sub-well, and resonances can be quasi-localized on these threes sub-wells.
The potential $V_\text{s}(z)$ approaches $0^-$ at infinity, therefore the scalar mode is not localized on the brane and does not contribute to the four-dimensional Newton potential.

\begin{figure}[!htb]
    \begin{center}
    \subfigure[$P_1$]{\label{figure P 21 }
        \includegraphics[width=4cm]{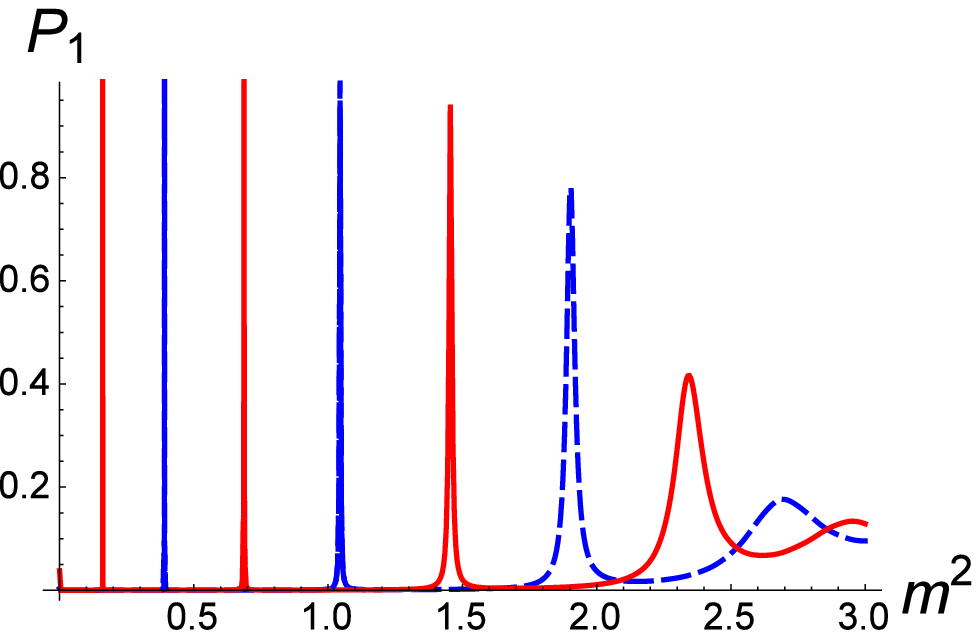}}
    \subfigure[$P_2$]{\label{figure P 22}
        \includegraphics[width=4cm]{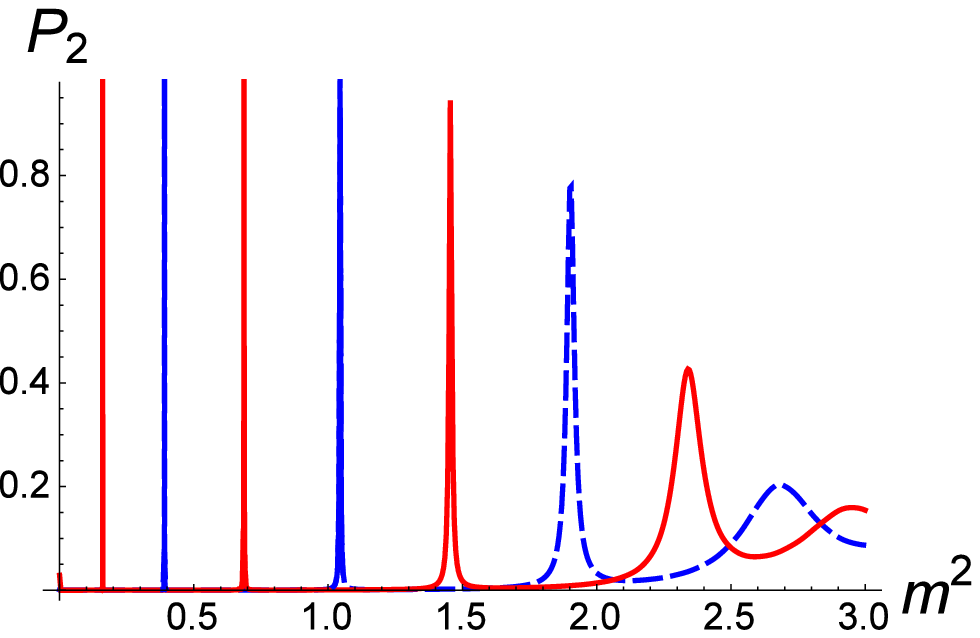}}
    \end{center}
    \caption{The relative probability  $P_1(m^2)$ and $P_2(m^2)$ of the even parity mode $t_\text{e}(z)$ (red solid lines) and the odd parity mode $t_\text{o}(z)$ (blue dashed lines) for the double brane model (\ref{WarpedFactorDoubleBrane}). The parameters are set to $k=1$
             and $b=1.8$. }  \label{figure 22}
    \end{figure}

\begin{figure}[!htb]
    \begin{center}
    \subfigure[$m^2=0.1606$]{\label{figure even 1}
        \includegraphics[width=4cm]{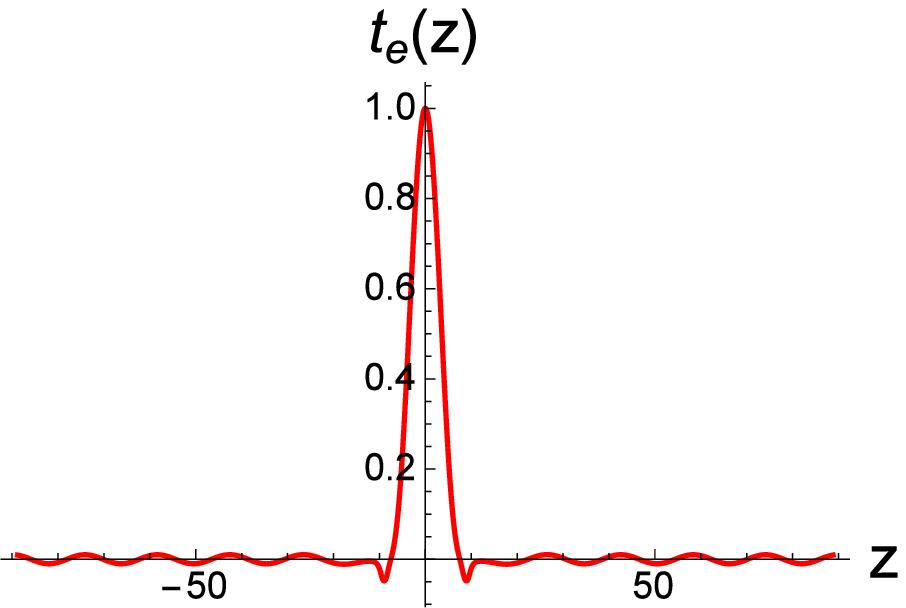}}
    \subfigure[$m^2=0.3907$]{\label{figure odd 1}
        \includegraphics[width=4cm]{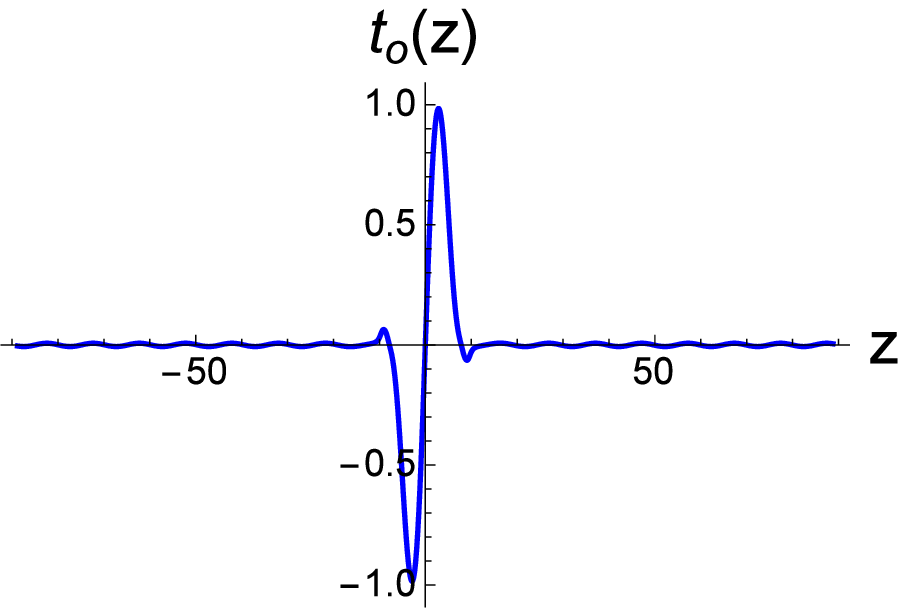}}
    \end{center}
    \caption{Plots of the first even and first odd parity resonance modes quasi-localized on the double brane described by (\ref{WarpedFactorDoubleBrane}). The parameters are set to $k=1$, $b=1.8$ and $m^2=0.1606$ for $t_\text{e} (z)$, $m^2=0.3907$ for $t_\text{o} (z)$.} \label{figure 23}
    \end{figure}

In this subsection, we would like to investigate the resonances quasi-localized on the double brane.
Since the effective potential has different structure, we introduce
two alternative definitions of the relative probability:
    \begin{eqnarray}
        \label{relative p1}
        P_1&=&\frac{\int^{z_m}_{-z_m}|t(z)|^2 dz}{\int^{10z_m}_{-10z_m}|t(z)|^2 dz},\\
        \label{relative p2}
        P_2&=&\frac{\int^{z_2}_{-z_2}|t(z)|^2 dz}{\int^{10z_2}_{-10z_2}|t(z)|^2 dz},
    \end{eqnarray}
where $z_m=\frac{z_1+z_2}{2}$, and ($z_1$, $z_2$) is the $z$-coordinate range of one of the sub-wells (see Fig.~\ref{figure potential 2}).
Following the same procedure, the relative probabilities $P_1$ and $P_2$ are plotted in
Fig.~\ref{figure 22}, and the spectra of the resonant modes calculated with the above two definitions of
relative probability are listed in Tab.~\ref{table p2}. From Tab.~\ref{table p2} we
can see that the even and odd parity modes appear alternately. Note that the
first even and odd resonance modes are not degenerate. For the two definitions of relative
probability, the difference between the spectra $\Delta m_n^2$ is much less than the mass square $m_n^2$. Thus, we may draw a conclusion that the two definitions give almost the same spectra of resonance modes. Table~\ref{table p2} shows that the FWHM increases with $m^2$, thus the life-time decreases with $m^2$. It can be seen that although there is more than one sub-well, the mass spectrum of the resonances is similar to the case of the single brane model in the last subsection. The wave functions of two resonances with mass square $m^2=0.1606$ and $m^2=0.3907$ are plotted in Fig.~\ref{figure 23}, which shows that the resonances are indeed quasi-localized on the double brane.

    \begin{table*}
        \centering
        \caption{The mass spectrum $m_{n}^2$, relative probability $P_{max}$, FWHM $\Gamma$ and life-time $\tau$ of the resonances quasi-localized on the double brane described by (\ref{WarpedFactorDoubleBrane2}).}
\begin{tabular}{|c|c|c|c|c|c|c|}
  \hline
  Parity & $m_{n}^2(P_1)$ &$m_{n}^2(P_2)$ & $\Delta m_{n}^2$ & $\Gamma$ & $\tau$&$P_{max}$ \\
  \hline
  even & $0.1606$ &$0.1605$ & $1\times10^{-4}$ & $1.248\times10^{-4}$ & $8015$&0.9978 \\
  \hline
   odd & $0.3907$ &$0.3908$ & $1\times10^{-4}$ & $3.760\times10^{-4}$ & $2660$&0.9982 \\
  \hline
  even & $0.6870$ &$0.6872$ & $2\times10^{-4}$ & $9.170\times10^{-4}$ & $1091$&0.9964 \\
  \hline
  odd & $1.0445$ &$1.0445$ & $0.0000$ & $2.153\times10^{-3}$ & $464.6$&0.9880 \\
  \hline
    even & $1.4555$ &$1.4565$ & $1.0\times10^{-3}$ & $9.597\times10^{-3}$ & $104.2$&0.02071 \\
  \hline
  odd & $1.9032$ &$1.9021$ & $1.1\times10^{-3}$ & $1.305\times10^{-2}$ & $76.64$&0.7916 \\
  \hline
      even & $2.3424$ &$2.3411$ & $1.3\times10^{-3}$ & $4.177\times10^{-2}$ & $23.94$&0.4164 \\
  \hline
  odd & $2.6949$ &$2.6802$ & $1.47\times10^{-2}$ & $0.1384$ & $7.223$&0.1762 \\
  \hline
\end{tabular}\label{table p2}
    \end{table*}

\subsection{Gravitational resonances quasi-localized on the sub-branes}\label{subSec33}

From Fig.~ \ref{figure potential 2} we can see that each sub-brane corresponds to a sub-well, which may support new kinds of gravitational resonances. Therefore, we investigate gravitational resonances quasi-localized on the sub-branes in this subsection.
We will analyze the influence of the distance between the two sub-branes and the thickness of the sub-branes. To this end, we consider the following warped factor
    \begin{eqnarray}
            \label{wf3}
          a(y) &= & \text{tanh}[k(y+d+b)]-\text{tanh}[k(y-d-b)] \nonumber \\
                &-&  \text{tanh}[k(y+d)]+\text{tanh}[k(y-d)], \label{WarpedFactorDoubleBrane2}
    \end{eqnarray}
where $2(b+d)$ is approximately the thickness of the brane,
and $2d$ is the distance between the two sub-branes in the physical coordinate $y$. The shapes of the warped factor
(\ref{WarpedFactorDoubleBrane2}) and the corresponding effective potential  $V_\text{t}(z)$ in this model are similar to those in subsection \ref{sbs 32}.
In order to investigate resonance modes which are only quasi-localized on the sub-branes, we define the corresponding relative
 probability $P_3$:
    \begin{eqnarray}
        \label{relative p3}
        P_3=
    \begin{cases}
    \frac{\int^{z_2}_{z_1}|t(z)|^2 dz}{\int^{z_m+5(z_2-z_1)}_{z_m-5(z_2-z_1)}|t(z)|^2 dz},& z_m\geq5(z_2-z_1)\\
    ~~\\
    \frac{\int^{z_2}_{z_1}|t(z)|^2 dz}{\int^{10(z_2-z_1)}_{0}|t(z)|^2 dz},& z_m<5(z_2-z_1)
    \end{cases}
    \end{eqnarray}
where $(z_1$,$z_2)$ is the $z$-coordinate range of one of the sub-wells (see Fig.~\ref{figure potential 2}
as a diagrammatic drawing),  and $z_m=\frac{z_1+z_2}{2}$.

\begin{figure}[!htb]
    \begin{center}
    \subfigure[$d=1.6$]{\label{figure P 31 }
        \includegraphics[width=4cm]{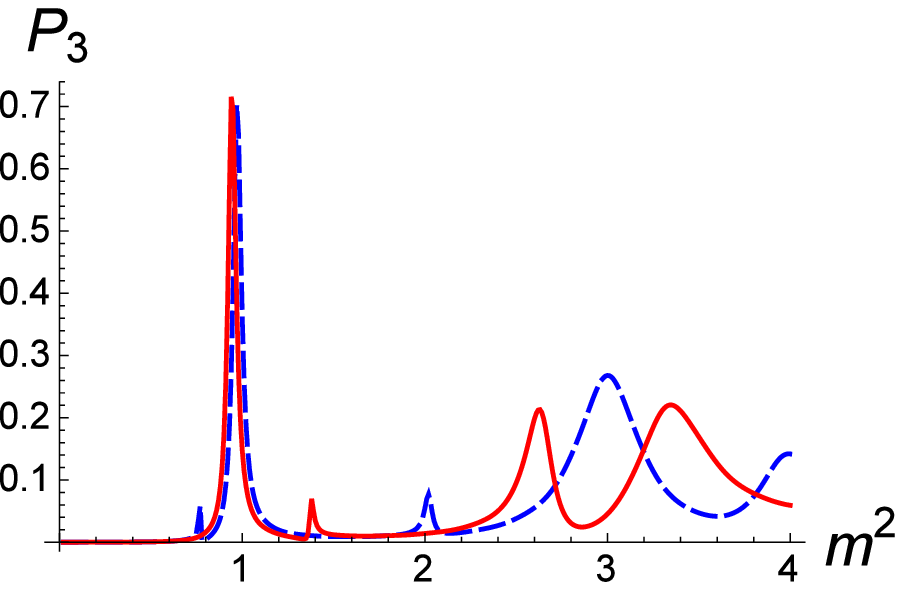}}
    \subfigure[$d=2.3$]{\label{figure P 32}
        \includegraphics[width=4cm]{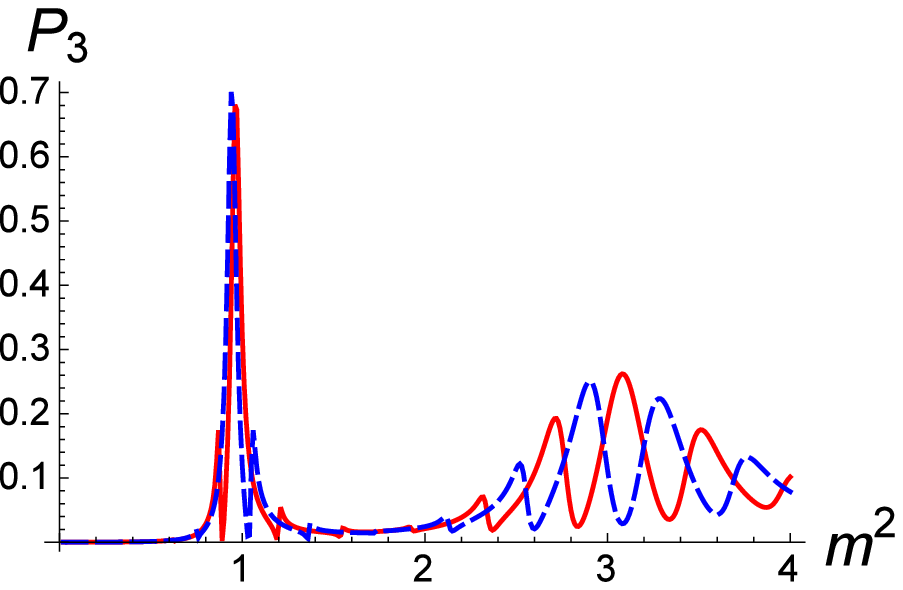}}
        \subfigure[$d=2.6$]{\label{figure P 33}
        \includegraphics[width=4cm]{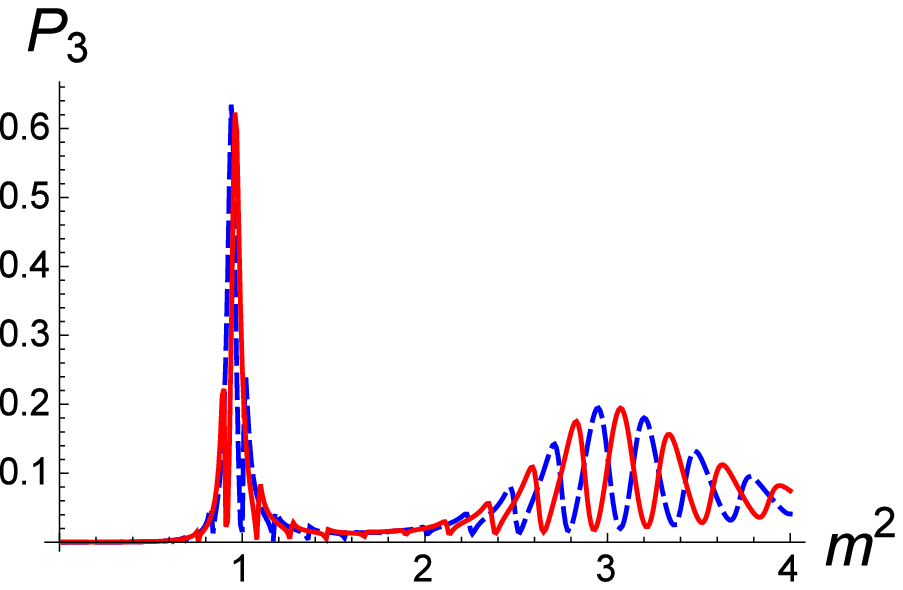}}
    \end{center}
    \caption{The relative  probability  $P_3(m^2)$ of the even parity mode $t_\text{e}(z)$ (red solid lines) and the odd parity mode $t_\text{o}(z)$ (blue dashed lines) quasi-localized on the sub-branes for the double brane model (\ref{WarpedFactorDoubleBrane2}). The parameters are set to $k=1$,
             and $b=7$. }  \label{figure 31}
    \end{figure}

\begin{figure}[!htb]
    \begin{center}
    \subfigure[$m^2=0.944$]{\label{figure even 311}
        \includegraphics[width=4cm]{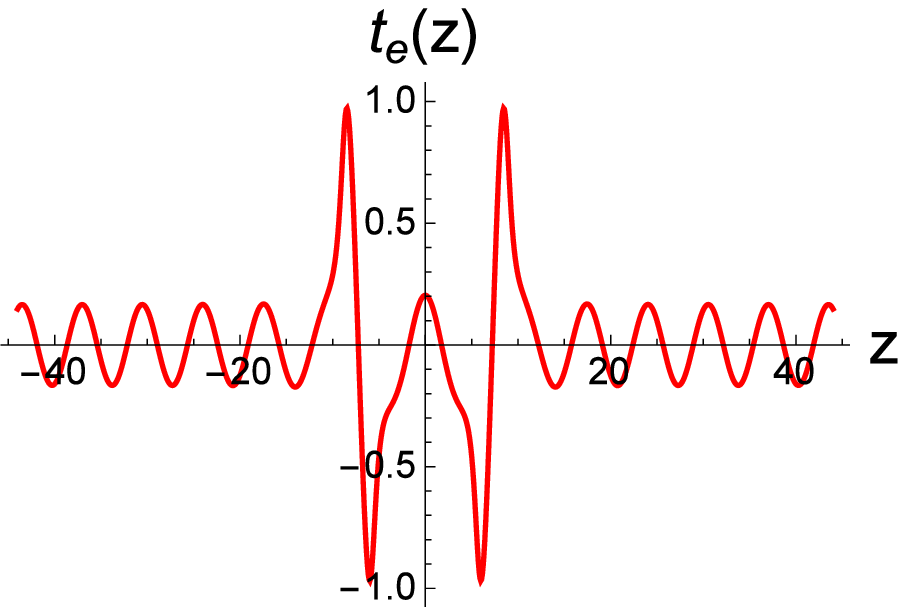}}
    \subfigure[$m^2=0.972$]{\label{figure odd 311}
        \includegraphics[width=4cm]{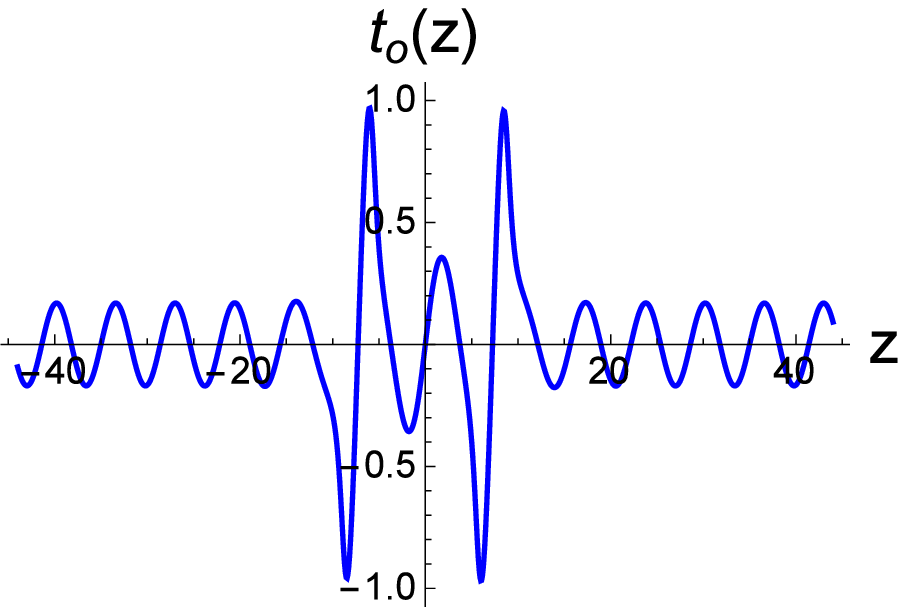}}
    \subfigure[$m^2=3.445$]{\label{figure even 312}
        \includegraphics[width=4cm ]{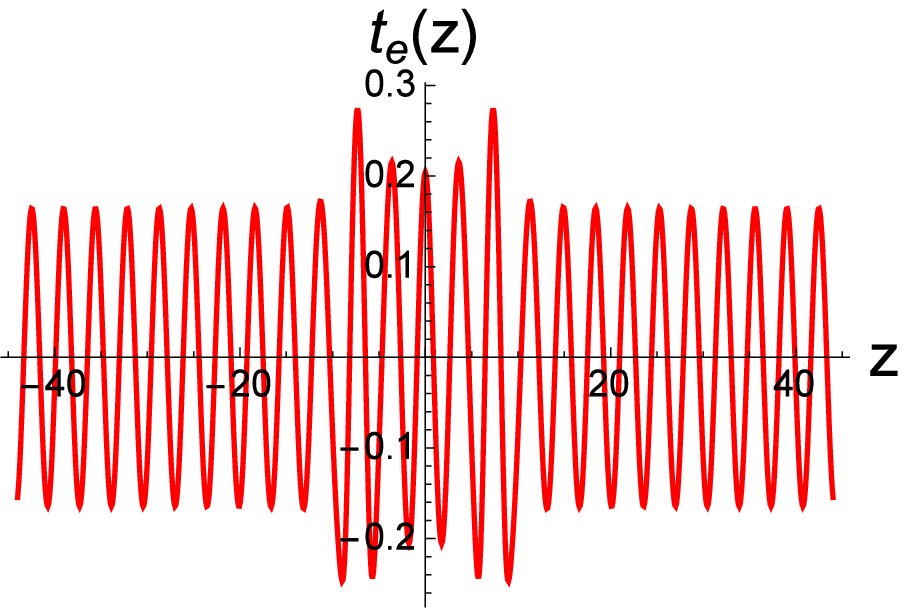}}
    \subfigure[$m^2=3.002$]{\label{figure odd 312}
        \includegraphics[width=4cm]{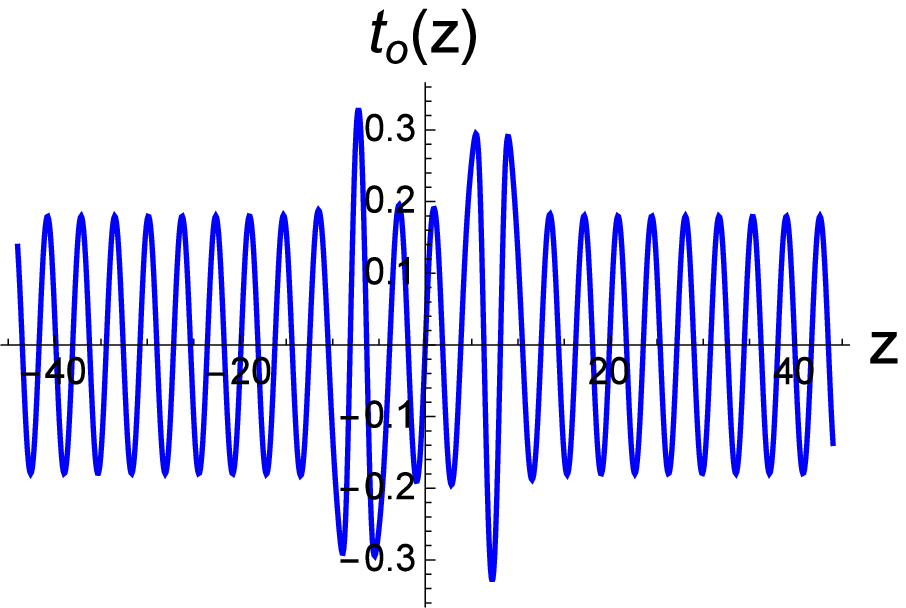}}
    \end{center}
    \caption{Plots of the even and odd parity resonance modes quasi-localized on the sub-branes for the double brane model (\ref{WarpedFactorDoubleBrane2}). The parameters are set to $k=1$, $b=7$, and $d=1.6$.} \label{figure wave 31}
    \end{figure}

\begin{figure}[!htb]
    \begin{center}
    \subfigure[$m^2=0.965$]{\label{figure even 321}
        \includegraphics[width=4cm]{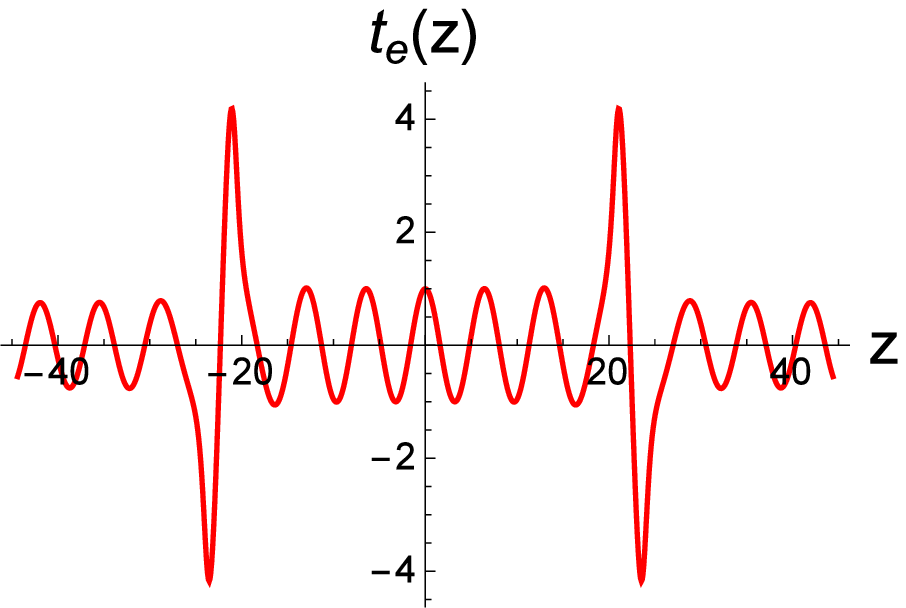}}
    \subfigure[$m^2=0.943$]{\label{figure odd 321}
        \includegraphics[width=4cm]{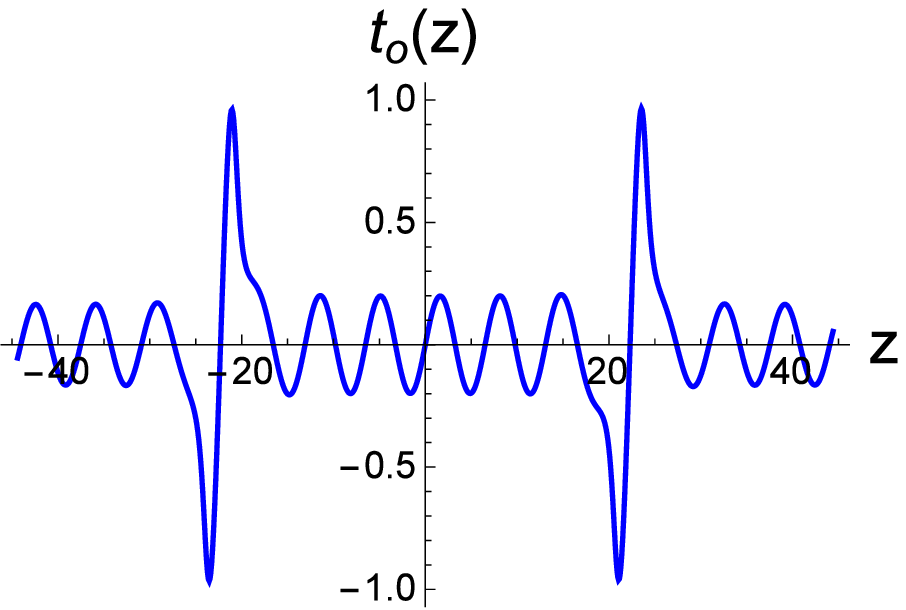}}
    \subfigure[$m^2=3.038$]{\label{figure even 322}
        \includegraphics[width=4cm ]{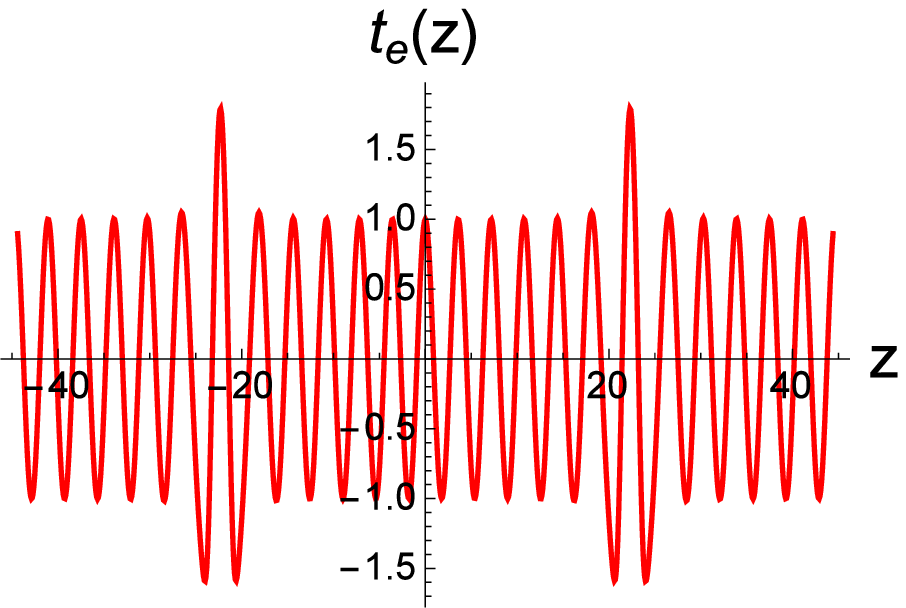}}
    \subfigure[$m^2=2.903$]{\label{figure odd 322}
        \includegraphics[width=4cm]{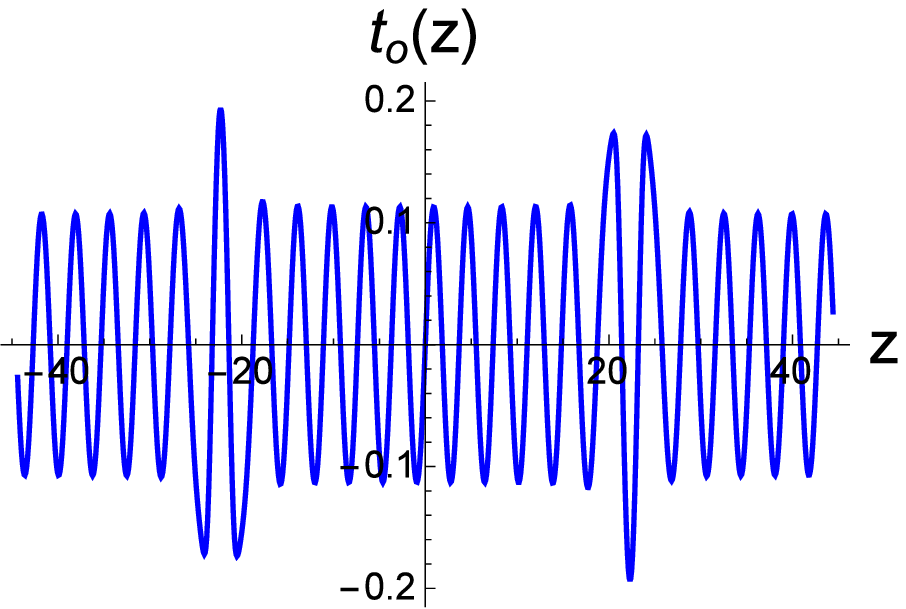}}
    \end{center}
    \caption{Plots of the even and odd parity resonance modes quasi-localized on the sub-branes for the double brane model (\ref{WarpedFactorDoubleBrane2}). The parameters are set to $k=1$, $b=7$, and $d=2.3$.} \label{figure wave 32}
    \end{figure}

\begin{figure}[!htb]
    \begin{center}
    \subfigure[$m^2=0.963$]{\label{figure even 331}
        \includegraphics[width=4cm]{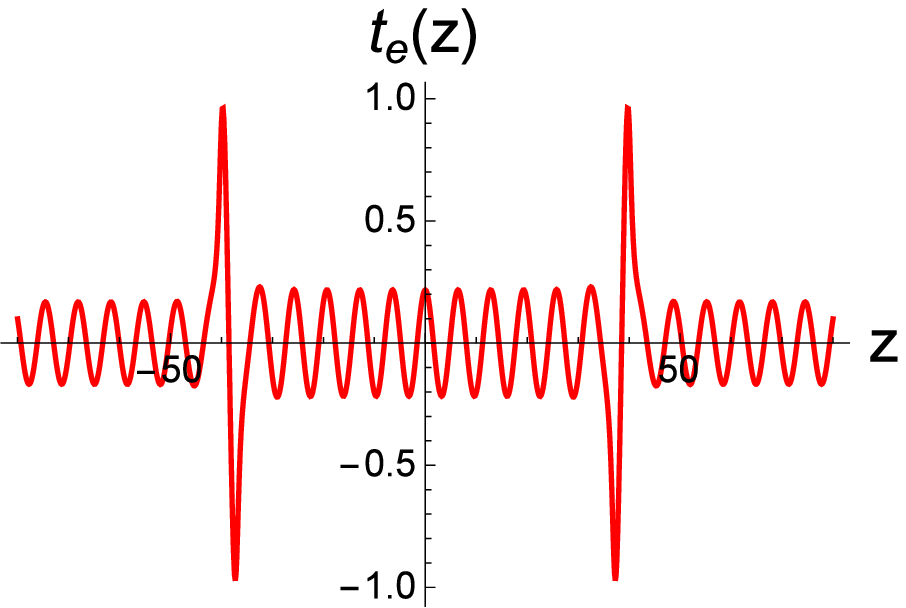}}
    \subfigure[$m^2=0.942$]{\label{figure odd 331}
        \includegraphics[width=4cm]{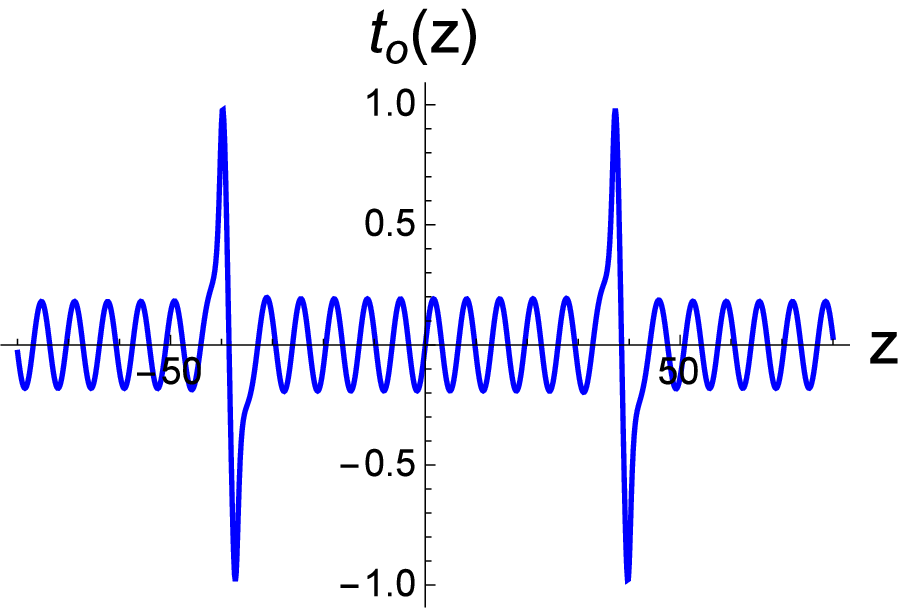}}
    \subfigure[$m^2=0.073$]{\label{figure even 332}
        \includegraphics[width=4cm ]{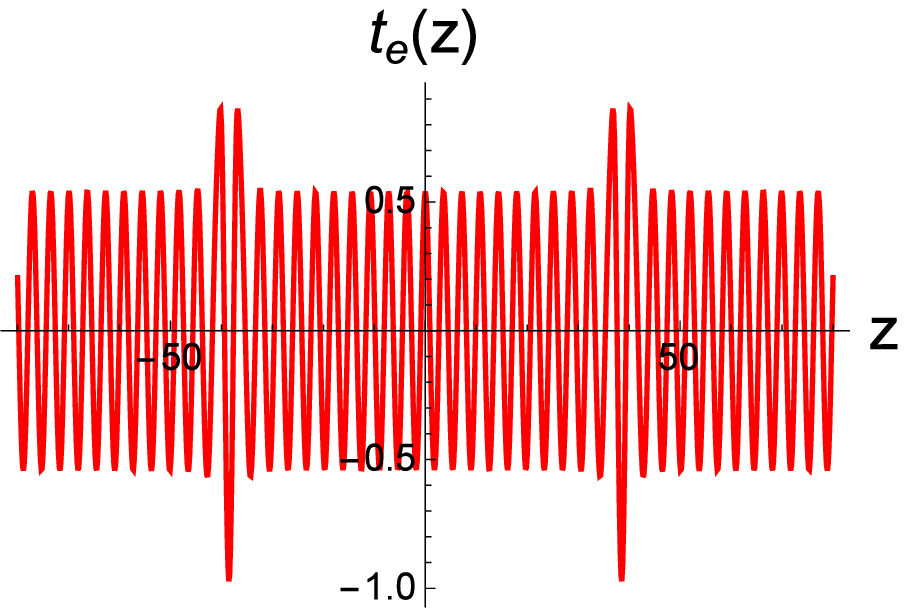}}
    \subfigure[$m^2=2.945$]{\label{figure odd 332}
        \includegraphics[width=4cm]{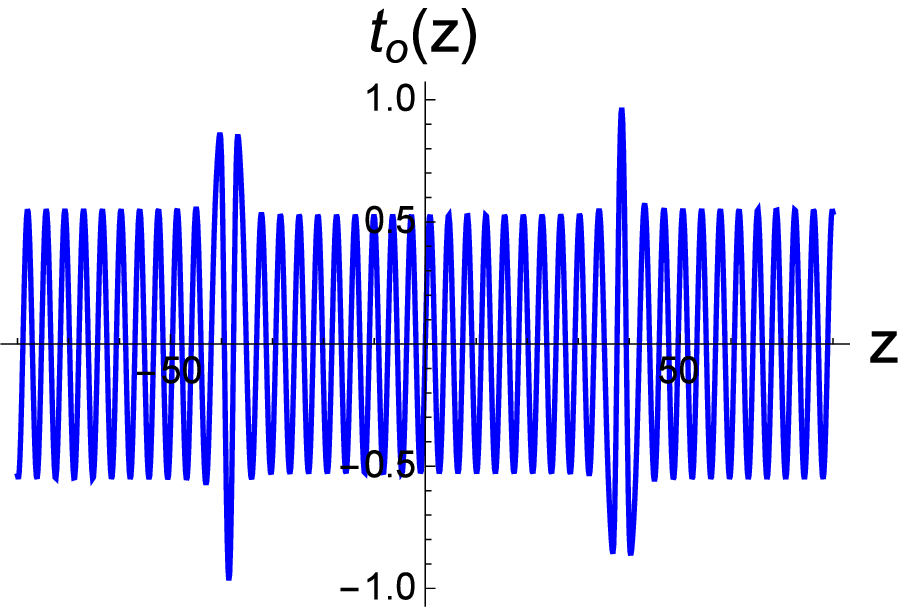}}
    \end{center}
    \caption{Plots of the even and odd parity resonance modes quasi-localized on the sub-branes for the double brane model (\ref{WarpedFactorDoubleBrane2}). The parameters are set to $k=1$, $b=7$, and $d=2.6$.} \label{figure wave 33}
    \end{figure}

Firstly, we fix the parameter $b$ and plot the shape of $P_3(m^2)$ for three values of the parameter $d$ in Fig.~\ref{figure 31}. The mass spectrum, relative probability, FLHM, and life-time of the resonances in two definitions of
relative probability are given in Tab.~\ref{table p3}.
Note that in Fig.~\ref{figure 31} and the figures of the relative probability in the following subsections, only the peaks which satisfy $P>0.1$ and have a FWHM represent the resonances. The relative probability of the resonance modes do not monotonically decrease with the mass
 square $m^2$, and there are only one odd and one even significant resonance modes, which is different from the case in single brane model. As the brane distance $d$ increases, more
 resonances appear. The mass spectra, FWHMs and life-times of a part of the resonances are shown in
 Tab.~\ref{table p3}. The wave functions $t_\text{e}(z)$ and $t_\text{o}(z)$ of a part of the resonance modes are shown in Figs.~\ref{figure wave 31}-\ref{figure wave 33}. For each resonance mode, the amplitude in the sub-wells is larger than the one out of the sub-wells, which shows that the definition~(\ref{relative p3}) of relative probability $P_3$ is proper. In Figs.~\ref{figure wave 31}-\ref{figure wave 33} the red and blue lines denote even and odd modes, respectively. Nevertheless, Figs.~\ref{figure even 312}, \ref{figure odd 312}, \ref{figure even 322}, \ref{figure odd 322}, \ref{figure even 332}, \ref{figure odd 332} are even modes with respect to the sub-branes, and the others are odd modes. This is crucial to the correction of the four-dimensional Newtonian potential, which will be demonstrated in the next section.

   \begin{table}
        \centering
        \caption{The mass spectrum $m_{n}^2$, relative probability $P_{max}$, FWHM $\Gamma$ and life-time $\tau$ of a part of the resonances quasi-localized on the sub-branes  with different brane distance $d$ for the double brane model (\ref{WarpedFactorDoubleBrane2}). }\label{table p3}
    \begin{tabular}{|c|c|c|c|c|c|}
  \hline
  $d$ &Parity & $m_{n}^2$ & $P_{max}$ & $\Gamma$ & $\tau$ \\
  \hline
  &even & $0.944$ &$0.7310$ & $0.02726$ & $36.68$ \\
  \cline{2-6}
    1.6&odd & $0.972$ &$0.7072$ & $0.02763$ & $36.20$ \\
  \cline{2-6}
  &odd & $3.002$ &$0.2680$ & $0.1227$ & $8.149$ \\
  \cline{2-6}
  &even & $3.345$ &$0.2206$ & $0.1385$ & $7.220$ \\
  \hline
  \hline
  &odd & $0.943$ &$0.7078$ & $0.03034$ & $32.96$ \\
  \cline{2-6}
    2.3&even & $0.965$ &$0.6955$ & $0.03076$ & $32.51$ \\
  \cline{2-6}
  &odd & $2.903$ &$0.2507$ & $0.06870$ & $14.56$ \\
  \cline{2-6}
  &even & $3.083$  & $0.2622$ & $0.07261$ & $13.77$ \\
  \hline

    \hline
  &even & $0.963$ &$0.6317$ & $0.02644$ & $37.82$ \\
  \cline{2-6}
    2.6&odd & $0.942$ &$0.6385$ & $0.02630$ & $38.02$ \\
  \cline{2-6}
  &odd & $2.945$ &$0.1947$ & $0.1877$ & $5.330$ \\
  \cline{2-6}
  &even & $3.073$  & $0.1945$ & $0.1863$ & $5.368$ \\
  \hline
\end{tabular}
    \end{table}

\begin{figure}[!htb]
    \begin{center}
    \subfigure[$b=5$]{\label{figure P 31 }
        \includegraphics[width=4cm]{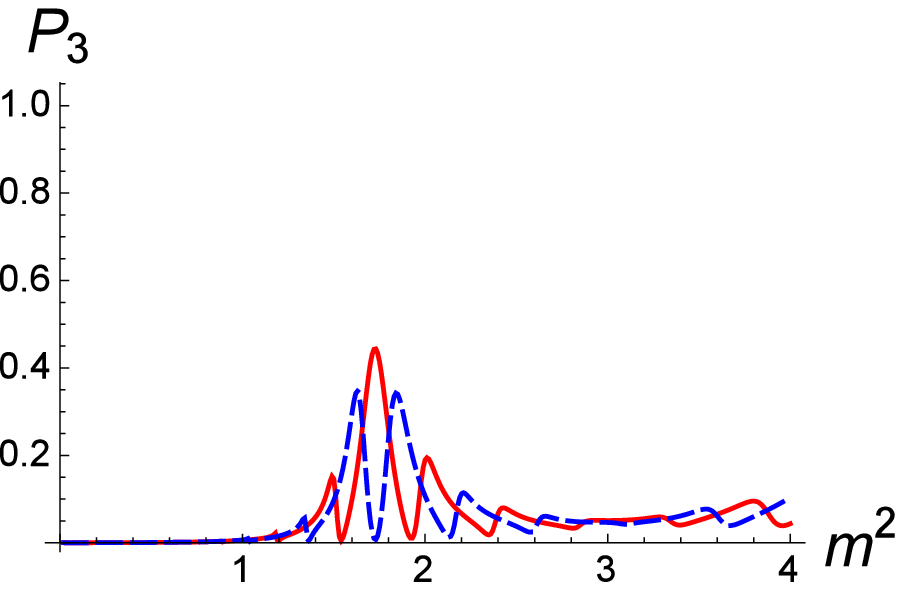}}
    \subfigure[$b=9$]{\label{figure P 32}
        \includegraphics[width=4cm]{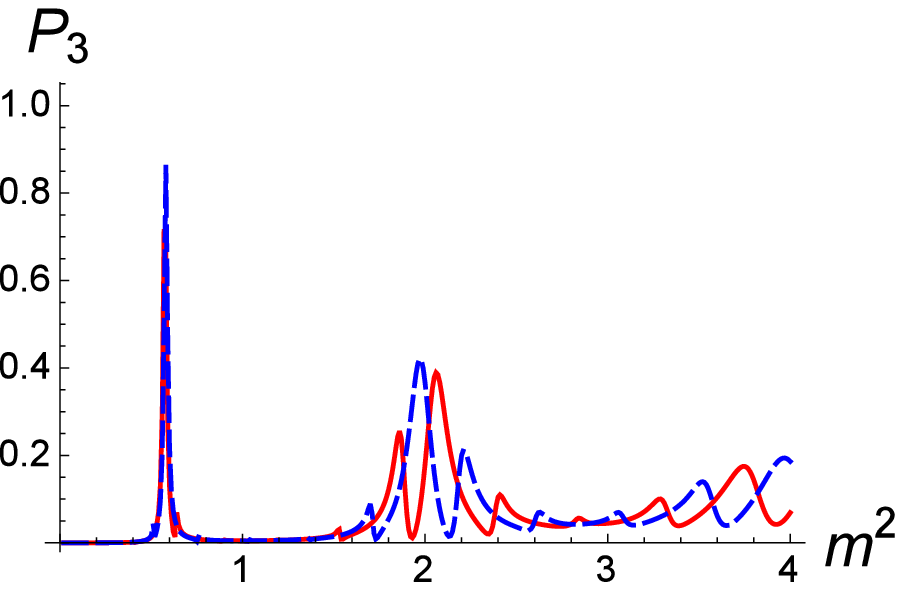}}
        \subfigure[$b=13$]{\label{figure P 33}
        \includegraphics[width=4cm]{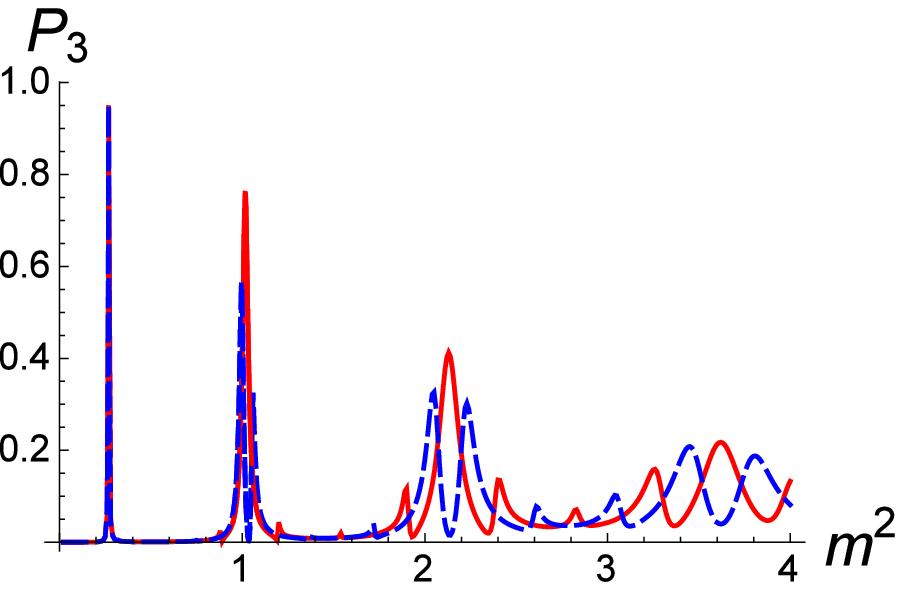}}
    \end{center}
    \caption{The relative probability  $P_3(m^2)$ of the even parity mode $t_\text{e}(z)$ (red solid lines) and the odd parity mode $t_\text{o}(z)$ (blue dashed lines) for the double brane model (\ref{WarpedFactorDoubleBrane2}). The parameters are set to $k=1$ and $d=2.3$. }  \label{figure 32}
    \end{figure}

Next, we analyze the influence of the thickness of sub-branes. We fix the parameter $d$ and plot
the shapes of the relative probability $P_3(m^2)$ for different values of the brane
thickness $b$ in Fig.~\ref{figure 32}. The mass spectrum, relative probability, FWHM and life-time of a part of
the resonances are given in Tab.~\ref{table p3 b2}.
It is shown that as the sub-brane thickness increases, the mass of the first even and odd modes decrease, while their relative probabilityincrease. Furthermore,
 for small sub-brane thickness $b=5$, there are only a group of resonances with small relative probability, while for large sub-brane thickness, the resonances with large relative probability appear. The wave functions of the resonances are similar to the ones in
 Fig. ~\ref{figure even 311}-\ref{figure odd 332}.

 Through the above demonstration, it can be seen that the character of the resonances quasi-localized on the sub-branes is quite different from that of the resonances quasi-localized on the
 double brane and single brane studied before.

   \begin{table}
        \centering
        \caption{The mass spectrum $m_{n}^2$, relative probability $P_{max}$, FWHM $\Gamma$ and life-time $\tau$ of a part of the resonances quasi-localized on sub-branes with different brane thickness $b$ for the double brane model (\ref{WarpedFactorDoubleBrane2}). }\label{table p3 b2}
    \begin{tabular}{|c|c|c|c|c|c|}
  \hline
  $b$ &Parity & $m_{n}^2$ & $P_{max}$ & $\Gamma$ & $\tau$ \\
  \hline
  &even &$1.726$& $0.444$ & $0.06316$ & $15.83$ \\
  \cline{2-6}
    5&odd &$1.628$& $0.347$ & $0.09062$ & $11.04$ \\
  \hline
  \hline
  &odd &$0.5744$& $0.835$ & $0.01445$ & $69.22$ \\
  \cline{2-6}
    9&even &$0.5794$ & $0.867$& $0.01504$ & $66.50$ \\
  \hline

    \hline
  &even &$0.2689$& $0.950$ & $0.004628$ & $216.1$ \\
  \cline{2-6}
    13&odd &$0.2683$& $0.952$ & $0.004633$ & $215.8$ \\
  \hline
\end{tabular}
    \end{table}

\subsection{Gravitational resonances quasi-localized between the sub-branes}\label{subSec34}

Figure~ \ref{figure potential 2} shows that the gravitational resonances could also be quasi-localized between the sub-branes, since sub-well between the sub-branes can support resonances. Therefore, we investigate the gravitational resonances quasi-localized between the sub-branes in this subsection. The warped factor $a(y)$ is also assumed as Eq.~(\ref{wf3}). The relative probability $P_4$ corresponding to the gravitational resonances quasi-localized between the two sub-branes is given by
    \begin{eqnarray}
        \label{relative p4}
        P_4&=&\frac{\int^{z_1}_{-z_1}|t(z)|^2 dz}{\int^{10z_1}_{-10z_1}|t(z)|^2 dz},
    \end{eqnarray}
where $z_1$ is shown in Fig.~\ref{figure potential 2}. Then we can analyze the influence of the thickness of sub-branes
and the distance between the two sub-branes, respectively.

\begin{figure}[!htb]
    \begin{center}
    \subfigure[$d=12$]{\label{figure P 31 }
        \includegraphics[width=4cm]{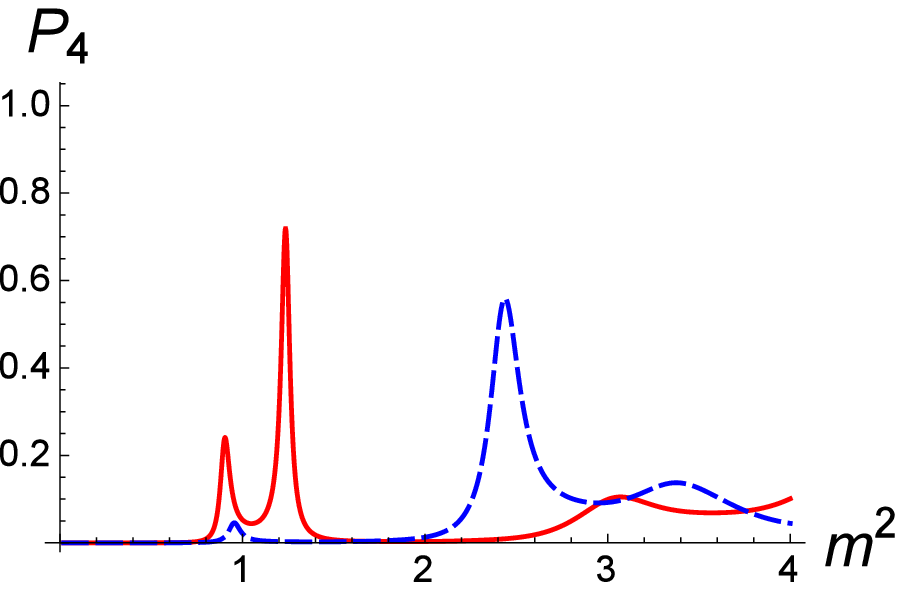}}
    \subfigure[$d=15$]{\label{figure P 32}
        \includegraphics[width=4cm]{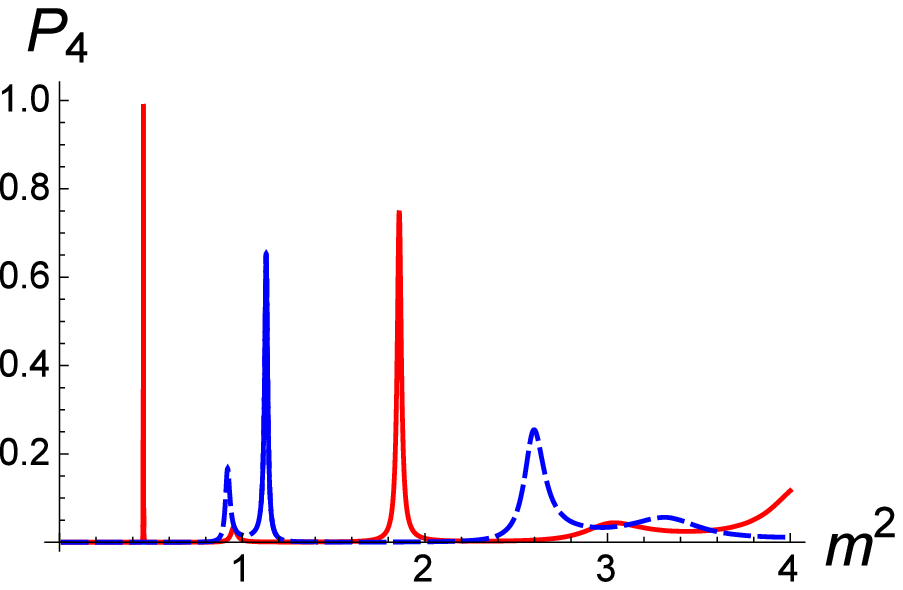}}
        \subfigure[$d=18$]{\label{figure P 33}
        \includegraphics[width=4cm]{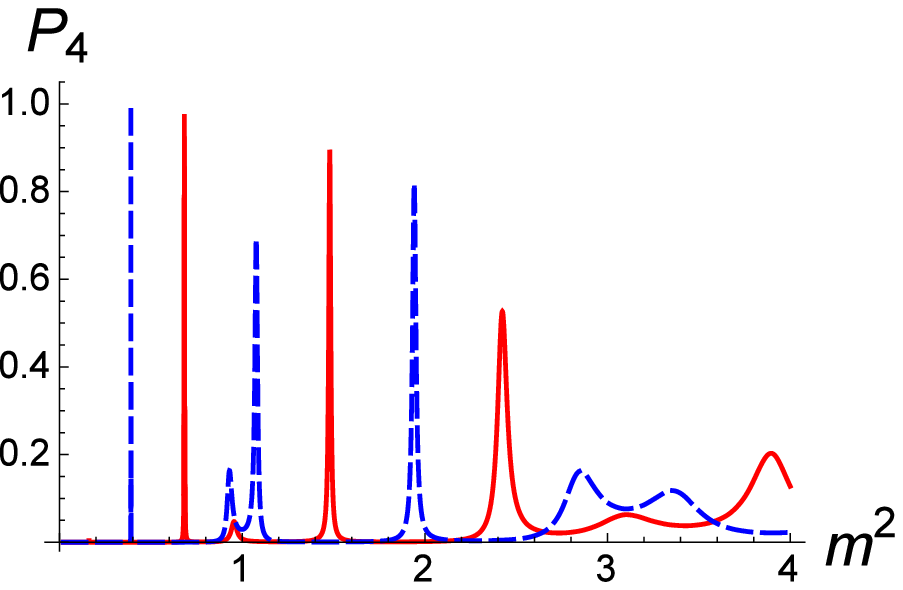}}
    \end{center}
    \caption{The relative  probability  $P_4(m^2)$ of the even parity mode $t_\text{e}(z)$ (red solid lines) and the odd parity mode $t_\text{o}(z)$ (blue dashed lines) quasi-localized between the sub-branes for different values of the distance $d$. The parameter $k$ is set to $k=1$, $b=7$, and $d=12,15,18$. }  \label{figure 41}
    \end{figure}

Firstly, we fix the sub-brane thickness and plot
the shapes of $P_3(m^2)$ for three values of the sub-brane
distance in Fig.~\ref{figure 41}, and list the information of some resonances in Tab.~\ref{table p4 thickness}. It can be seen that the number and life-time of the resonances increase with the width $d$ of the middle sub-well, which is similar to the case of a single brane \cite{Csaki:2000fc,Xie:2013rka}. While the relative probability of the resonances does not monotonically decrease with the mass
 square $m^2$, which is very different from the case of a single brane \cite{Csaki:2000fc,Xie:2013rka}.

   \begin{table}
        \centering
        \caption{The mass spectrum $m_{n}^2$, relative probability $P_{max}$, FWHM $\Gamma$ and life-time $\tau$ of a part of the resonances quasi-localized between the sub-branes with different brane distance for the double brane model (\ref{WarpedFactorDoubleBrane2}). }\label{table p4 thickness}
    \begin{tabular}{|c|c|c|c|c|c|}
  \hline
  $d$ &Parity & $m_{n}^2$ & $P_{max}$ & $\Gamma$ & $\tau$ \\
  \hline
  &even & $1.235$ &$0.7185$ & $0.02655$ & $37.66$ \\
  \cline{2-6}
    1.2&odd & $2.438$ &$0.5590$ & $0.03195$ & $31.30$ \\
  \hline
  \hline
  &odd & $0.459$ &$0.990$ & $0.000594$ & $1693$ \\
  \cline{2-6}
    1.5&even & $1.131$ &$0.653$ & $0.009404$ & $106.3$ \\
  \hline

    \hline
  &even & $0.683$ &$0.9765$ & $0.001452$ & $688.8$ \\
  \cline{2-6}
    1.8&odd & $0.391$ &$0.9979$ & $0.0003200$ & $3125$ \\
  \hline
\end{tabular}
    \end{table}

\begin{figure}[!htb]
    \begin{center}
    \subfigure[$b=11$]{\label{figure P 31 }
        \includegraphics[width=4cm]{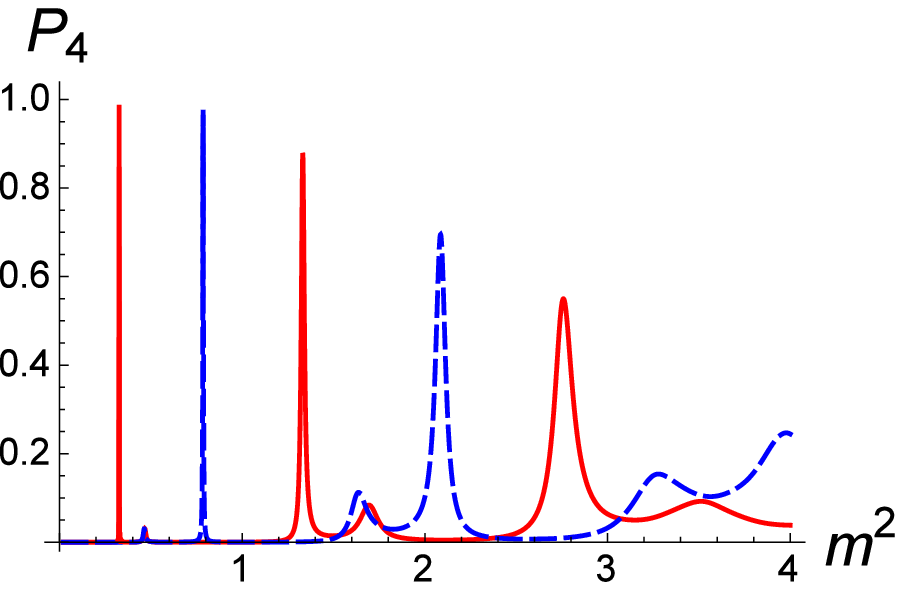}}
    \subfigure[$b=11.9$]{\label{figure P 32}
        \includegraphics[width=4cm]{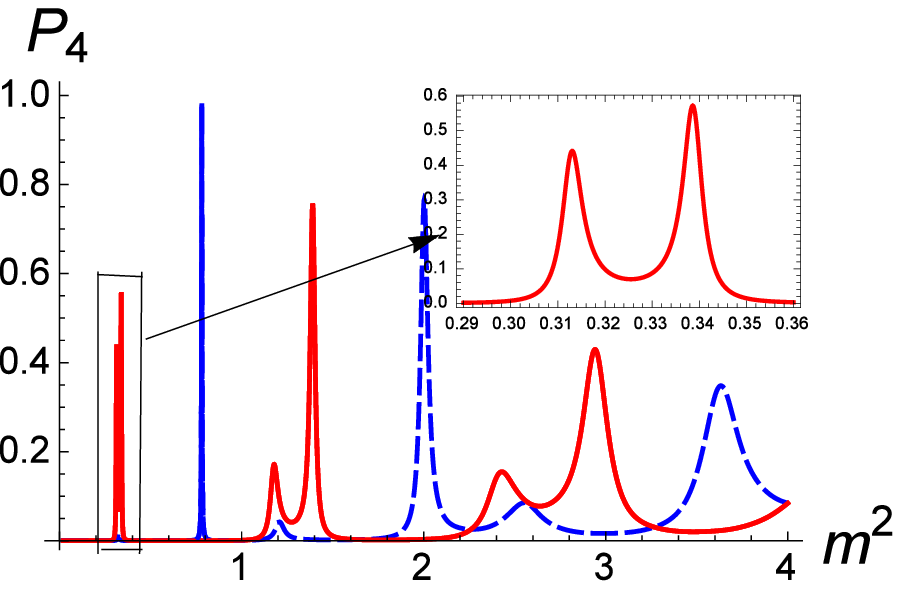}}
    \subfigure[$b=14.9$]{\label{figure P 33}
        \includegraphics[width=4cm]{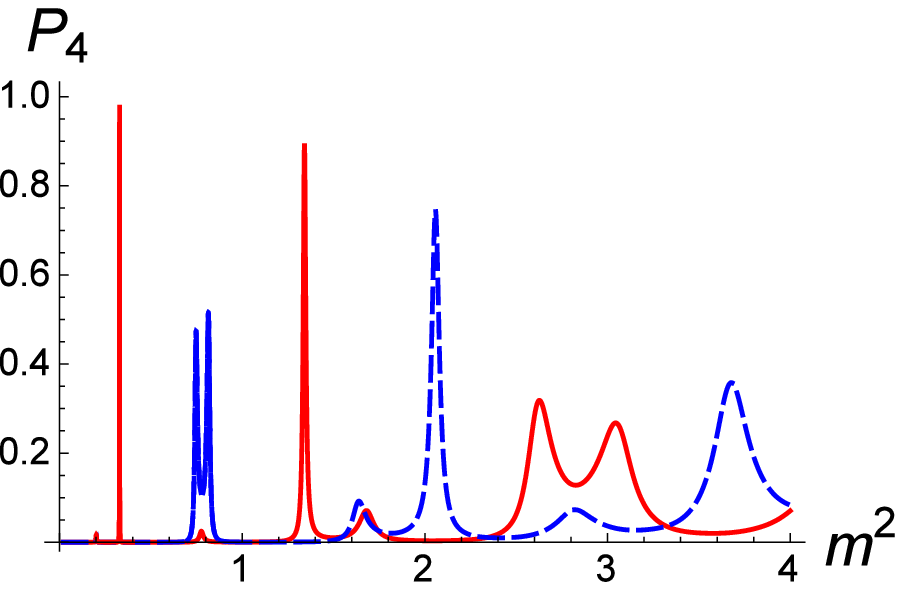}}
    \subfigure[$b=16$]{\label{figure P 33}
        \includegraphics[width=4cm]{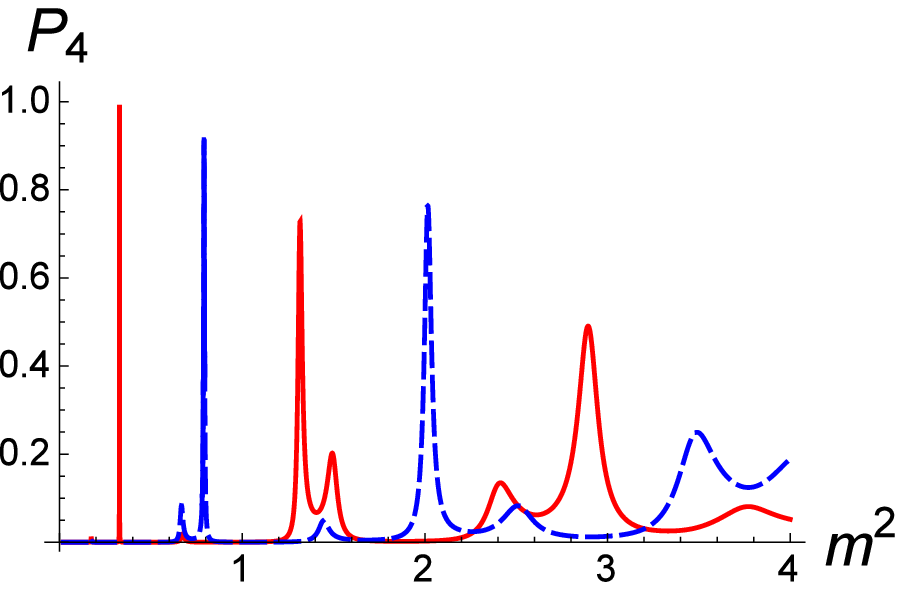}}
    \end{center}
    \caption{The relative probability  $P_4(m^2)$ of the even parity mode $t_\text{e}(z)$ (red solid lines) and the odd parity mode $t_\text{o}(z)$ (blue dashed lines) quasi-localized between the sub-branes for different values of the brane thickness $b$. The parameters are set to $k=1$, $d=2.3$, and $b=11,11.9,14.9,16$. }  \label{figure 42}
    \end{figure}

Next, we fix the distance $d$ and plot
the shapes of $P_4(m^2)$ for different values of the brane
thickness $b$ in Fig.~\ref{figure 42}. It is shown that though the parameter $b$ is related to the sub-wells on the left and right rather than the one in the middle, it has an important impact on the resonance quasi-localized
between the sub-branes. With the increasing of the parameter $b$, an even or odd resonance splits into two resonances with small relative probability, and then the two resonances become a single resonance again. While the other resonances are not significantly changed with the parameter $b$, which is different from all the cases above or the case in a single brane model.

\section{Conclusion and discussion}\label{Sec4}
In this work, we discussed gravitational resonances quasi-localized on different locations of mimetic branes.
Firstly we considered a single brane as an example and used the relative probability in Ref.~\cite{Liu:2009ve} to investigate the gravitational resonances. Then we considered a double brane and investigated resonances quasi-localized on the double brane, on the sub-branes and between the sub-branes, respectively. For the first case, since the effective potential splits into two sub-wells, we introduced two alternative definitions of the  relative probability. In each definition, we obtained the spectrum of gravitational resonances and showed
that the two spectra are almost the same. We also obtained the FWHM and life-time of the resonances.
For the second case, we introduced another definition of
relative probability to investigate the gravitational resonances quasi-localized on the sub-branes.
The influence of the distance between the two sub-branes and the thickness of the sub-branes was analyzed. We found that more gravitational resonance modes appear with the increasing of the distance between the two sub-branes or the thickness of the sub-branes. Especially, we found some new feature of the resonances.
For the third case, we investigated the resonances quasi-localized between the two sub-branes.
We showed that the number and life-time of the resonances increase with the distance between the two sub-branes.
As the brane thickness increases, an even or odd
resonance will split into two resonances with smaller relative probability, and then the two resonances will become a single resonance again.

Finally, we will discuss briefly the contribution of the resonances to the four-dimensional Newtonian potential. For the case that the brane locates at $z=z_0$ along the extra dimension, the correction from resonances to the four-dimensional Newtonian potential of a mass point $M$ is given by \cite{Csaki:2000fc,Lykken:1999nb}
    \begin{eqnarray}
        \label{correction potential}
         U(r) \sim G_N \frac{M}{r}
         \left[1+ \int^{\infty}_{0}dm ~ \text{e}^{-mr} t_m^2(z_0) \right],
    \end{eqnarray}
where $r$ is the distance to the mass point on the brane, and $t_m(z)$ is the normalized wave function of the resonance with mass $m$ and the normalization constant $N_m$ is related to $m$. For subsections~\ref{subSec31}, \ref{subSec32}, and \ref{subSec34}, $z_0=0$, and so only the even modes contribute to the four-dimensional Newtonian potential as $t_\text{o}(0)=0$. For subsection~\ref{subSec33}, the resonances are quasi-localized on the sub-branes, therefore $z_0=\int^{b+d}_0 \frac{dy}{a(y)} $, and the resonances shown in Figs.~\ref{figure even 312}, \ref{figure odd 312}, \ref{figure even 322}, \ref{figure odd 322}, \ref{figure even 332}, \ref{figure odd 332} contribute to the four-dimensional Newtonian potential.
 The resonances which have larger life-time escape to the five-dimension more slowly \cite{Csaki:2000pp}. On the other hand, it is shown in section \ref{Sec3} that these resonances have larger $t_m(z_0)$. Therefor, the resonances which have larger life-time contribute to the Newtonian potential more than other resonances.
The mass spectrum of the resonances show that the mass scale $m$ of the resonances is of about $m\sim k$, where $k$ is the scale on the brane. Usually we assume that $k$ is much less than the Planck scale, therefore $m$ is also much less than the Planck mass.
According to the analysis of Refs.~\cite{Csaki:2000fc,Lykken:1999nb}, the normalization constant $N_m$ is decided by the asymptotic behavior of the effective potential $V_\text{t}(z)$ in (\ref{PotentialV}) at the boundary of the extra dimension. For the asymptotic AdS$_5$ solutions (\ref{WarpedFactorSingleBrane}), (\ref{WarpedFactorDoubleBrane}), and (\ref{WarpedFactorDoubleBrane2}) considered in this paper, we have $V_\text{t}(z) \propto \frac{15}{4z^2}$ at $|z|\gg 1/k$ and $t_m(0)\sim (m/k)^{1/2}$ for the brane located at $z_0=0$, for which the correction to the four-dimensional Newtonian potential is
    \begin{eqnarray}
        \label{correction potential}
         U(r) \sim G_N \frac{M}{r}
         \left[1+ \frac{C}{(kr)^2} \right],
    \end{eqnarray}
where $C$ is a dimensionless constant determined by the structure of the brane.  For the case of the double brane, the correction also occurs at the scale of $r\sim 1/k$, but it has a complex form because of the rich structure of the mass spectrum of the resonance KK modes.
According the recent test of the gravitational inverse-square law, the usual Newtonian potential still holds down to a length scale at 59$\mu$m \cite{TanLuoPRL2016}. Therefore, the thickness of the brane, $1/k$, should be much less than 59$\mu$m.

\section*{Acknowledgement}
This work was supported by the National Natural Science Foundation of China (Grants No. 11875151 and No. 11522541) and the Fundamental Research Funds for the Central Universities (Grants No.	531107051196, No. lzujbky-2018-k11, and No. lzujbky-2017-it68).


\begin{thebibliography}{9}
\bibitem{ArkaniHamed:1998rs}
  N.~Arkani-Hamed, S.~Dimopoulos, and G.~R.~Dvali,
  {\em The Hierarchy problem and new dimensions at a millimeter},
  Phys. Lett. {\bf B 429} (1998) 263
  [arXiv:hep-ph/9803315].

\bibitem{Randall:1999ee}
  L.~Randall and R.~Sundrum,
  {\em A Large mass hierarchy from a small extra dimension},
  Phys.\ Rev.\ Lett.\  {\bf 83} (1999) 3370
  [arXiv:hep-ph/9905221].

\bibitem{Randall:1999vf}
  L.~Randall and R.~Sundrum,
  {\em An Alternative to compactification},
  Phys. Rev. Lett.  {\bf 83} (1999) 4690
  [arXiv:hep-th/9906064].

\bibitem{Kim:2000mc}
  J.~E.~Kim, B.~Kyae, and H.~M.~Lee,
  {\em Randall-Sundrum model for selftuning the cosmological constant},
  Phys. Rev. Lett.  {\bf 86} (2001) 4223
  [arXiv:hep-th/0011118].
\bibitem{Kobayashi:2001jd}
  S.~Kobayashi, K.~Koyama, and J.~Soda,
  {\em Thick brane worlds and their stability},
  Phys. Rev. {\bf D 65} (2002) 064014
  [arXiv:hep-th/0107025].

\bibitem{Giovannini:2001fh}
  M.~Giovannini,
  {\em Gauge invariant fluctuations of scalar branes},
  Phys. Rev. {\bf D 64} (2001) 064023
  [arXiv:hep-th/0106041].

\bibitem{Bazeia:2003aw}
  D.~Bazeia, C.~Furtado, and A.~R.~Gomes,
  {\em Brane structure from scalar field in warped space-time},
  JCAP {\bf 0402} (2004) 002
  [arXiv:hep-th/0308034].

\bibitem{Bazeia:2004dh}
  D.~Bazeia and A.~R.~Gomes,
  {\em Bloch brane},
  JHEP {\bf 0405} (2004) 012
  [arXiv:hep-th/0403141].

\bibitem{Csaki:2000fc}
  C.~Csaki, J.~Erlich, T.~J.~Hollowood, and Y.~Shirman,
  {\em Universal aspects of gravity localized on thick branes},
  Nucl.\ Phys.\ {\bf B 581} (2000) 309
  [arXiv:hep-th/0001033].

\bibitem{DeWolfe:1999cp}
  O.~DeWolfe, D.~Z.~Freedman, S.~S.~Gubser, and A.~Karch,
  {\em Modeling the fifth-dimension with scalars and gravity},
  Phys. Rev. {\bf D 62} (2000) 046008
  [arXiv:hep-th/9909134].

\bibitem{Gremm:1999pj}
  M.~Gremm,
  {\em Four-dimensional gravity on a thick domain wall},
  Phys. Lett. {\bf B 478} (2000) 434
  [arXiv:hep-th/9912060].

\bibitem{Gremm:2000dj}
  M.~Gremm,
  {\em Thick domain walls and singular spaces},
  Phys. Rev. {\bf D 62} (2000) 044017
  [arXiv:hep-th/0002040].

\bibitem{Afonso:2006gi}
  V.~I.~Afonso, D.~Bazeia, and L.~Losano,
  {\em First-order formalism for bent brane},
  Phys. Lett. {\bf B 634} (2006) 526
  [arXiv:hep-th/0601069].

\bibitem{Bazeia:2002sd}
  D.~Bazeia, L.~Losano, and C.~Wotzasek,
  {\em Domain walls in three field models},
  Phys. Rev. {\bf D 66} (2002) 105025
  [arXiv:hep-th/0206031].

\bibitem{Dzhunushaliev:2008gk}
  V.~Dzhunushaliev, V.~Folomeev, S.~Myrzakul, and R.~Myrzakulov,
  {\em Phantom thick brane in 5D bulk},
  Mod.\ Phys.\ Lett.\ {\bf A 23} (2008) 2811
  [arXiv:0804.0151 [gr-qc]].

\bibitem{Neupane:2010ey}
  I.~P.~Neupane,
  {\em De Sitter brane-world, localization of gravity, and the cosmological constant},
  Phys. Rev. {\bf D 83} (2011) 086004
  [arXiv:1011.6357 [hep-th]].

\bibitem{Bajc:1999mh}
  B.~Bajc and G.~Gabadadze,
  {\em Localization of matter and cosmological constant on a brane in anti-de Sitter space},
  Phys. Lett. {\bf B 474} (2000) 282
  [arXiv:hep-th/9912232].

\bibitem{Ringeval:2001cq}
  C.~Ringeval, P.~Peter, and J.~P.~Uzan,
  {\em Localization of massive fermions on the brane},
  Phys. Rev. {\bf D 65} (2002) 044016
  [arXiv:hep-th/0109194].

\bibitem{Bagger:2004rr}
  J.~A.~Bagger and D.~V.~Belyaev,
  {\em Brane-localized Goldstone fermions in bulk supergravity},
  Phys. Rev. {\bf D 72} (2005) 065007
  [arXiv:hep-th/0406126].

\bibitem{Liu:2007ku}
  Y.~X.~Liu, X.~H.~Zhang, L.~D.~Zhang, and Y.~S.~Duan,
  {\em Localization of Matters on Pure Geometrical Thick Branes},
  JHEP {\bf 0802} (2008) 067
  [arXiv:0708.0065 [hep-th]].

\bibitem{Liu:2008wd}
  Y.~X.~Liu, L.~D.~Zhang, S.~W.~Wei, and Y.~S.~Duan,
  {\em Localization and Mass Spectrum of Matters on Weyl Thick Branes},
  JHEP {\bf 0808} (2008) 041
  [arXiv:0803.0098 [hep-th]].

\bibitem{Ghoroku:2001zu}
  K.~Ghoroku and A.~Nakamura,
  {\em Massive vector trapping as a gauge boson on a brane},
  Phys. Rev. {\bf D 65} (2002) 084017
  [arXiv:hep-th/0106145].

\bibitem{Kehagias:2000au}
  A.~Kehagias and K.~Tamvakis,
  {\em Localized gravitons, gauge bosons and chiral fermions in smooth spaces generated by a bounce},
  Phys. Lett. {\bf B 504} (2001) 38
  [hep-th/0010112].


\bibitem{Cruz:2013uwa}
  W.~T.~Cruz, L.~J.~S.~Sousa, R.~V.~Maluf, and C.~A.~S.~Almeida,
  {\em Graviton resonances on two-field thick branes},
  Phys. Lett. {\bf B 730} (2014) 314
  [arXiv:1310.4085 [hep-th]].

\bibitem{Xie:2013rka}
  Q.~Y.~Xie, J.~Yang, and L.~Zhao,
  {\em Resonance Mass Spectra of Gravity and Fermion on Bloch Branes},
  Phys. Rev. {\bf D 88} (2013) 105014
  [arXiv:1310.4585 [hep-th]].

\bibitem{Xu:2014jda}
  Z.~G.~Xu, Y.~Zhong, H.~Yu, and Y.~X.~Liu,
  {\em The structure of $f(R)$-brane model},
  Eur.\ Phys.\ J.\ {\bf C 75} (2015) 368
  [arXiv:1405.6277 [hep-th]].

\bibitem{Csaki:2000pp}
  C.~Csaki, J.~Erlich, and T.~J.~Hollowood,
  {\em Quasilocalization of gravity by resonant modes},
  Phys. Rev. Lett.  {\bf 84} (2000) 5932
  [arXiv:hep-th/0002161].

\bibitem{Yu:2015wma}
  H.~Yu, Y.~Zhong, B.~M.~Gu, and Y.~X.~Liu,
  {\em Gravitational resonances on $f(R)$-brane},
  Eur.\ Phys.\ J.\ {\bf C 76} (2016) 195
  [arXiv:1506.06458 [gr-qc]].



\bibitem{Nollert:1999ji}
  H.~P.~Nollert,
  {\em TOPICAL REVIEW: Quasinormal modes: the characteristic `sound' of black holes and neutron stars},
  Class.\ Quant.\ Grav.\  {\bf 16} (1999) R159.


\bibitem{Liu:2017gcn}
 Y. X. Liu, {\em Introduction to Extra Dimensions and Thick Braneworlds},
  arXiv:1707.08541
  [M. L. Ge, R. G. Cai, and Y. X. Liu, \emph{Memorial Volume for Yi-Shi Duan}, World Scientific 211-275 (2018)].

\bibitem{Sotiriou:2008rp}
  T.~P.~Sotiriou and V.~Faraoni,
  {\em f(R) Theories Of Gravity},
  Rev.\ Mod.\ Phys.\  {\bf 82} (2010) 451
  [arXiv:0805.1726 [gr-qc]].


\bibitem{Nojiri:2006ri}
  S.~Nojiri and S.~D.~Odintsov,
  {\em Introduction to modified gravity and gravitational alternative for dark energy},
  eConf {\bf C 0602061} (2006) 06
   [Int.\ J.\ Geom.\ Meth.\ Mod.\ Phys.\  {\bf 4} (2007) 115]
  [arXiv:hep-th/0601213].

\bibitem{Nojiri:2017ncd}
  S.~Nojiri, S.~D.~Odintsov and V.~K.~Oikonomou,
  {\em Modified Gravity Theories on a Nutshell: Inflation, Bounce and Late-time Evolution},
  Phys.\ Rept.\  {\bf 692} (2017) 1
  [arXiv:1705.11098 [gr-qc]].


\bibitem{Kastor:2017knv}
  D.~Kastor, S.~Ray, and J.~Traschen,
  {\em Lovelock Branes},
  Class.\ Quant.\ Grav.\  {\bf 34} (2017) 195005
  [arXiv:1706.06684 [gr-qc]].

\bibitem{Bazeia:2017mjd}
  D.~Bazeia, M.~A.~Marques, and R.~Menezes,
  {\em Generalized Born-Infeld-like models for kinks and branes},
  EPL {\bf 118} (2017) 11001
  [arXiv:1703.05848 [hep-th]].

\bibitem{Afonso:2007zz}
  V.~I.~Afonso, D.~Bazeia, R.~Menezes, and A.~Y.~Petrov,
  {\em f(R)-Brane},
  Phys. Lett. {\bf B 658} (2007) 71
  [arXiv:0710.3790 [hep-th]].

\bibitem{Guo:2014bxa}
  B.~Guo, Y.~X.~Liu, and K.~Yang,
  {\em Brane worlds in gravity with auxiliary fields},
  Eur.\ Phys.\ J.\ {\bf C 75} (2015) 63
  [arXiv:1405.0074 [hep-th]].

\bibitem{German:2013sk}
  G.~German, A.~Herrera--Aguilar, D.~Malagon--Morejon, I.~Quiros, and R.~da Rocha,
  {\em Study of field fluctuations and their localization in a thick braneworld generated by gravity nonminimally coupled to a scalar field with the Gauss-Bonnet term},
  Phys. Rev. {\bf D 89} (2014) 026004
  [arXiv:1301.6444 [hep-th]].

\bibitem{Arias:2002ew}
  O.~Arias, R.~Cardenas, and I.~Quiros,
  {\em Thick brane worlds arising from pure geometry},
  Nucl.\ Phys.\ {\bf B 643} (2002) 187
  [arXiv:hep-th/0202130].

\bibitem{Zhong:2015pta}
  Y.~Zhong and Y.~X.~Liu,
  {\em Pure geometric thick $f(R)$-branes: stability and localization of gravity},
  Eur.\ Phys.\ J.\ {\bf C 76} (2016) 321
  [arXiv:1507.00630 [hep-th]].

\bibitem{Yang:2012hu}
  J.~Yang, Y.~L.~Li, Y.~Zhong, and Y.~Li,
  {\em Thick Brane Split Caused by Spacetime Torsion},
  Phys. Rev. {\bf D 85} (2012) 084033
  [arXiv:1202.0129 [hep-th]].

\bibitem{daSilva:2017jbx}
  P.~M.~L.~T.~da Silva and J.~M.~Hoff da Silva,
  {\em f(R)-Einstein-Palatini formalism and smooth branes},
  Eur.\ Phys.\ J.\ Plus {\bf 132} (2017) 437.

\bibitem{Yang:2017evd}
  K.~Yang, W.~D.~Guo, Z.~C.~Lin, and Y.~X.~Liu,
  {\em Domain wall brane in a reduced Born-Infeld-$f(T)$ theory},
  Phys. Lett. {\bf B 782} (2018) 170
  [arXiv:1709.01047 [hep-th]].

\bibitem{Guo:2015qbt}
  W.~D.~Guo, Q.~M.~Fu, Y.~P.~Zhang, and Y.~X.~Liu,
  {\em Tensor perturbations of $f(T)$-branes},
  Phys. Rev. {\bf D 93} (2016) 044002
  [arXiv:1511.07143 [hep-th]].

\bibitem{Karam:2018squ}
  A.~Karam, A.~Lykkas and K.~Tamvakis,
  {\em Frame-invariant approach to higher-dimensional scalar-tensor gravity},
  Phys. Rev. {\bf D 97} (2018) 124036
  [arXiv:1803.04960 [gr-qc]].

\bibitem{Guo:2018tpo}
  W.~D.~Guo, Y.~Zhong, K.~Yang, T.~T.~Sui, and Y.~X.~Liu,
  {\em Thick brane in mimetic $f(T)$ gravity},
  [arXiv:1805.05650 [hep-th]].


\bibitem{Chamseddine:2013kea}
  A.~H.~Chamseddine and V.~Mukhanov,
  {\em Mimetic Dark Matter},
  JHEP {\bf 1311} (2013) 135
  [arXiv:1308.5410 [astro-ph.CO]].

\bibitem{Chamseddine:2014vna}
  A.~H.~Chamseddine, V.~Mukhanov, and A.~Vikman,
 {\em Cosmology with Mimetic Matter},
  JCAP {\bf 1406} (2014) 017
  [arXiv:1403.3961 [astro-ph.CO]].

\bibitem{Dutta:2017fjw}
  J.~Dutta, W.~Khyllep, E.~N.~Saridakis, N.~Tamanini, and S.~Vagnozzi,
  {\em Cosmological dynamics of mimetic gravity},
  [arXiv:1711.07290 [gr-qc]].

\bibitem{Matsumoto:2015wja}
  J.~Matsumoto, S.~D.~Odintsov, and S.~V.~Sushkov,
  {\em Cosmological perturbations in a mimetic matter model}
  Phys. Rev. {\bf D 91} (2015) 064062
  [arXiv:1501.02149 [gr-qc]].

\bibitem{Myrzakulov:2015kda}
  R.~Myrzakulov, L.~Sebastiani, S.~Vagnozzi, and S.~Zerbini,
  {\em Static spherically symmetric solutions in mimetic gravity: rotation curves and wormholes},
  Class.\ Quant.\ Grav.\  {\bf 33} (2016) 125005
  [arXiv:1510.02284 [gr-qc]].

\bibitem{Vagnozzi:2017ilo}
  S.~Vagnozzi,
  {\em Recovering a MOND-like acceleration law in mimetic gravity},
  Class.\ Quant.\ Grav.\  {\bf 34} (2017) 185006
  [arXiv:1708.00603 [gr-qc]].

\bibitem{Nojiri:2014zqa}
  S.~Nojiri and S.~D.~Odintsov,
  {\em Mimetic $F(R)$ gravity: inflation, dark energy and bounce},
  Mod.\ Phys.\ Lett.\ {\bf A 29} (2014) 1450211
  [arXiv:1408.3561 [hep-th]].


\bibitem{Astashenok2015}
  A.~V.~Astashenok, S.~D.~Odintsov and V.~K.~Oikonomou,
  {\em Modified Gauss-Bonnet gravity with the Lagrange multiplier constraint as mimetic theory},
  Class.\ Quant.\ Grav.\  {\bf 32} (2015) 185007
  [arXiv:1504.04861 [gr-qc]].




\bibitem{Odintsov:2015ocy}
  S.~D.~Odintsov and V.~K.~Oikonomou,
  {\em Mimetic $F(R)$ inflation confronted with Planck and BICEP2/Keck Array data},
  Astrophys.\ Space Sci.\  {\bf 361} (2016) 174
  [arXiv:1512.09275 [gr-qc]].



\bibitem{Odintsov:2016imq}
  S.~D.~Odintsov and V.~K.~Oikonomou,
  {\em Unimodular Mimetic $F(R)$ Inflation},
  Astrophys.\ Space Sci.\  {\bf 361} (2016) 236
  [arXiv:1602.05645 [gr-qc]].

\bibitem{Odintsov:2016oyz}
  S.~D.~Odintsov and V.~K.~Oikonomou,
  {\em Dark Energy Oscillations in Mimetic $F(R)$ Gravity},
  Phys. Rev. {\bf D 94} (2016) 044012
  [arXiv:1608.00165 [gr-qc]].


\bibitem{Nojiri:2016vhu}
  S.~Nojiri, S.~D.~Odintsov and V.~K.~Oikonomou,
  {\em Viable Mimetic Completion of Unified Inflation-Dark Energy Evolution in Modified Gravity},
  Phys. Rev. {\bf D 94} (2016) 104050
  [arXiv:1608.07806 [gr-qc]].



\bibitem{Odintsov:2018ggm}
  S.~D.~Odintsov and V.~K.~Oikonomou,
  {\em The reconstruction of $f(\phi)R$ and mimetic gravity from viable slow-roll inflation},
  Nucl. Phys. {\bf B 929} (2018) 79
  [arXiv:1801.10529 [gr-qc]].

\bibitem{Sadeghnezhad:2017hmr}
  N.~Sadeghnezhad and K.~Nozari,
  {\em Braneworld Mimetic Cosmology},
  Phys. Lett. {\bf B 769} (2017) 134
  [arXiv:1703.06269 [gr-qc]].


\bibitem{Zhong:2017uhn}
  Y.~Zhong, Y.~Zhong, Y.~P.~Zhang, and Y.~X.~Liu,
  {\em Thick branes with inner structure in mimetic gravity},
  Eur. Phys. J. {\bf C 78} (2018) 45
  [arXiv:1711.09413 [hep-th]].


\bibitem{Liu:2009ve}
  Y.~X.~Liu, J.~Yang, Z.~H.~Zhao, C.~E.~Fu, and Y.~S.~Duan,
  {\em Fermion Localization and Resonances on A de Sitter Thick Brane},
  Phys. Rev. {\bf D 80} (2009) 065019,
  [arXiv:0904.1785 [hep-th]].

\bibitem{Almeida:2009jc}
  C.~A.~S.~Almeida, M.~M.~Ferreira, Jr., A.~R.~Gomes, and R.~Casana,
  {\em Fermion localization and resonances on two-field thick branes},
  Phys. Rev. {\bf D 79} (2009) 125022
  [arXiv:0901.3543 [hep-th]].

\bibitem{DuZhao2013}
 Y.-Z. Du, L. Zhao, Y. Zhong, C.-E Fu, and H. Guo,
 \em{Resonances of Kalb-Ramond field on symmetric and asymmetric thick branes},
 Phys. Rev. {\bf D 88} (2013) 024009,
 arXiv:1301.3204[hep-th].

\bibitem{Landim1105.5573}
 R. R. Landim, G. Alencar, M. O. Tahim, and R. N. Costa Filho,
 \em{A transfer matrix method for resonances in randall-sundrum models},
 JHEP {\bf 1108} (2011) 071, [arXiv:1105.5573].

%

\bibitem{Gregory:2000}
  R.~Gregory, V.~A.~Rubakov, and S.~M.~Sibiryakov,
  {\em Opening up extra dimensions at ultra large scales},
  Phys. Rev. Lett.  {\bf 84} (2000) 5928
  [arXiv:hep-th/0002072].




\bibitem{Lykken:1999nb}
  J.~D.~Lykken and L.~Randall,
  {\em The Shape of gravity}
  JHEP {\bf 0006} (2000) 014
  [arXiv:hep-th/9908076].
%


\bibitem{TanLuoPRL2016}
  W.-H. Tan et al.,
  {\em New test of the gravitational inverse-square law at the submillimeter range with dual modulation and
compensation},
Phys. Rev. Lett. {\bf 116} (2016) 131101.



\end{thebibliography}
\end{document}